\documentclass[aps,superscriptaddress,preprintnumbers,floatfix,nofootinbib,titlepage]{revtex4}
\usepackage{graphicx}
\usepackage{epsfig}
\newcommand{\be}{\begin{eqnarray}}
\newcommand{\ee}{\end{eqnarray}}
\newcommand\del{\partial}
\def\conj#1{{{#1}^{*}}}

\newcommand{\mat}{\left ( \begin{array}{cc}}
\newcommand{\emat}{\end{array} \right )}
\newcommand{\matf}{\left ( +
\begin{array}{cccc}}
\newcommand{\ematf}{\end{array} \right )}            
\newcommand{\matt}{\left \begin{array}{ccc}}
\newcommand{\ematt}{\end{array} \right )}
\newcommand{\vect}{\left ( \begin{array}{c}}
\newcommand{\evect}{\end{array} \right )}

\newcommand{\nn}{\nonumber } 
\newcommand{\tK}{Z^{n=1}_\nu}
\newcommand{\tKz}{Z^{n=1}_{\nu=0}}
\newcommand{\nK}{K_{N+n}}
\newcommand{\np}{p_{N+n}}
\def\d{\partial}
\newcommand{\tr}{{\rm Tr}}

\begin{document}
\setcounter{page}{0} 
\thispagestyle{empty}

%\wideabs{

%\begin{flushleft}
%AOSV-v10.tex
%\end{flushleft}

%\begin{flushright}
%SUNY-NTG-04/03 \\
%NORDITA-2004-78 HE 
%\end{flushright}

\preprint{SUNY-NTG-04/03}
\preprint{NORDITA-2004-78 HE}

\title{Unquenched QCD Dirac Operator Spectra at Nonzero
Baryon Chemical Potential
}

\author{G. Akemann}
\affiliation{
Department of Mathematical Sciences,
Brunel University West London, 
Uxbridge UB8 3PH, United Kingdom}
\author{J.C. Osborn}
\affiliation{Physics Department, Boston University,
Boston, MA 02215, USA}
\author{K. Splittorff}
\affiliation{Nordita, Blegdamsvej 17, DK-2100, Copenhagen {\O}, Denmark}
\author{J.J.M. Verbaarschot}
\affiliation{Department of Physics and Astronomy, SUNY, Stony Brook,
 New York 11794, USA}

%\author {G. Akemann$^1$, J.C. Osborn$^2$, K. Splittorff$^3$ and
%  J.J.M. Verbaarschot$^4$}
%\address{$^1$ Department of Mathematical Sciences,
%Brunel University West London,\\
%Uxbridge UB8 3PH, United Kingdom \\
%$^2$ Physics Department, Boston University, Boston, MA 02215, USA \\
%$^3$ Nordita, Blegdamsvej 17, DK-2100, Copenhagen {\O}, Denmark \\ 
%$^4$ Department of Physics and Astronomy, SUNY, Stony Brook, 
%New York 11794, USA} 

\date   {\today}
%\maketitle

\begin  {abstract}

The microscopic spectral density of the 
QCD Dirac operator at nonzero baryon chemical potential for an arbitrary 
number of quark flavors was derived recently from a random matrix model
with the global symmetries of QCD. In this paper we show that these
results and extensions thereof
can be obtained from the replica limit of a Toda lattice equation. This
naturally leads to a factorized form into bosonic and fermionic
QCD-like partition functions. In the microscopic limit these partition
functions are given by the static limit of a chiral Lagrangian that follows
from the symmetry breaking pattern.
In particular, we elucidate the role of the
singularity of the bosonic partition function in the orthogonal 
polynomials approach. A detailed discussion of 
the spectral density for one and two flavors is given.
\end {abstract}

\maketitle

%}

%\thispagestyle{empty}

%\newpage
%\tableofcontents
%\widetext
%\onecolumn

%%%%%%%%%%%%%%%%%%%%%%%%%%%%%%%%%%%%%%%%%%%%%%%%%%%%%%%%%%%%%%%%%%%%%%%%%%%%%%

\section{Introduction} 

One of the features that makes QCD at nonzero baryon chemical potential 
both elusive and interesting is that the Euclidean Dirac operator does not have
any Hermiticity properties. The non-Hermiticity  occurs in an 
essential way because a finite baryon number density is obtained by
promoting the propagation of quarks in the forward time direction and
inhibiting the propagation in the backward time direction. Because of
the non-Hermiticity the quark determinant attains a complex phase, which
prevents the analysis of the QCD partition function by means of
probabilistic methods. An approximation that is often used at zero
chemical potential is to ignore the fermion determinant altogether.
However, this approximation fails dramatically at nonzero chemical
potential \cite{Barbour}. In particular, the critical chemical potential
at zero temperature
is determined by the pion mass instead of the nucleon mass. 
Therefore, the main source of information
for QCD at nonzero baryon density is based on simplified models or on
perturbative expansions at very high densities which may never
be accessed experimentally. Lattice QCD simulations are feasable
in a region where the ratio of the chemical potential and the temperature
is sufficiently small so that extrapolation from an imaginary chemical
potential \cite{owe,maria} or from zero chemial potential
\cite{allton} becomes reliable (for a critical review of
the field and additional references, we refer to \cite{frit}).

In this paper, we will investigate an observable
for which nonperturbative results can be obtained for unquenched
QCD at nonzero baryon density. This observable is the spectral density
of the QCD Dirac operator. At nonzero chemical potential its support 
is a two-dimensional domain. 
The naive argument that observables do not
depend on the chemical potential at zero temperature does not
apply to the spectral density of the Euclidean Dirac operator.
This puzzling fact can be understood as follows. To define an
eigenvalue density we need both the eigenvalues and the complex
conjugate eigenvalues or the Dirac operator and its complex
conjugate. However, complex conjugation is equivalent to changing
the sign of the chemical potential. Therefore, the generating function
for the QCD Dirac spectrum has to contain 
both quarks and conjugate quarks in the valence sector
that have an opposite baryon charge.
This opens the possibility that the
low energy limit of this theory contains Goldstone bosons made out of
quarks and conjugate anti-quarks which have a nonzero baryon number
\cite{Gocksch,misha}. More formally, the chemical potential in the
generating function
breaks
the flavor symmetry resulting in a chiral Lagrangian that becomes
dependent on the chemical potential.
The region of the Dirac spectrum we wish to analyze
determines the mass of the valence quarks \cite{SV,Vplb}.
In order to determine the eigenvalue density of the Dirac operator at 
small eigenvalues the masses of the valence quarks are equally small. 
Consequently, the mass of the charged Goldstone modes is small and the 
generating function depends on the chemical potential even at small
values of $\mu$.    
 
The low energy effective theory for the generating
function of the Dirac spectrum is a theory of
Goldstone modes determined by spontaneous
breaking of chiral symmetry. This effective theory is a version of a chiral
Lagrangian \cite{GL} which takes into account the presence of 
the conjugate
quarks. The allowed terms in the effective theory are determined by the flavor
symmetries and the way they are broken by 
the quark masses and the chemical potential:
the effective theory must break the flavor symmetries in 
precisely the 
same way as in the generating function we started from.
To be specific, the Goldstone field, $U$, 
takes values  in the coset $U\in
SU(N_f+2n)$, where $n$ is the number of additional quark -- conjugate quark
pairs. 
The leading order chiral Lagrangian in the standard
counting scheme is the usual non-linear sigma model \cite{Weinberg}. 
The dependence of the chemical
potential is completely fixed by the flavor symmetries. It enters as the
zeroth component of an external vector field through a shift of the Euclidean
time derivative (see e.g. \cite{GL,KST,KSTVZ,dominique-JV,eff}) 
\be    
\d_0U \to \d_0U - i \mu [B,U],  
\label{gauge}
\ee
where $\mu$ is the chemical potential and 
$B$ is the charge matrix. Since the baryon charge of 
conjugate quarks is opposite
to the standard quarks \cite{Gocksch} the charge matrix $B$ is 
not proportional to the identity and the commutator is non-vanishing.

Although the eigenvalue spectrum of the Dirac operator is 
not a direct physical
observable, we hope that detailed knowledge of this observable in a
nonperturbative domain where no other information is available, will 
ultimately lead to a better understanding of the problems that
hinder numerical simulations at nonzero baryon chemical potential. 
At zero temperature these problems are manifest
unless $\mu^2F_\pi^2V\ll 1$, where $F_\pi$ is the pion
decay constant. In this paper, we will 
focus on a scaling regime where the product
$\mu^2F_\pi^2V$ is fixed as the volume, $V$, is taken to infinity. 
The fixed combination $\mu^2F_\pi^2V$ can take any value and our 
results will show the effect of the fermion determinant on the
Dirac spectrum. 
The quark masses
will be taken such that $m_\pi$ scales like $\mu$ with 
$m_\pi^2 F_\pi^2 V$ kept fixed. 
To be specific, we consider the range \cite{LS,SV}
\be
m_\pi ,\, \mu \ll \frac{1}{L} \ll \Lambda \ ,
\label{epslim}
\ee
as the volume $L^4=V$ is taken to infinity, and $\Lambda$ is the scale of the 
lightest non-Goldstone particle.
This regime is sometimes known as the ergodic domain or as the 
epsilon regime\footnote{The epsilon regime is the regime where 
$m_\pi \sim O(\epsilon^2)$ and $1/L \sim O(\epsilon)$. These conditions
are more strict than the inequality (\ref{epslim}). For example, we
are still in the ergodic domain if $1/L \sim O(\epsilon^{3/2})$.} 
of chiral perturbation theory \cite{GLeps}.  
In this domain the zero modes of the
Goldstone fields dominate the partition function which reduces  
to a static integral over the Goldstone manifold \cite{GLeps,dominique-JV}.
The static integral is completely determined by the flavor symmetries of 
the QCD partition function. This implies that any theory with the same
flavor symmetries and flavor symmetry breaking is described 
by the same static integral. The only memory of the underlying 
theory is two coupling constants, namely the chiral condensate $\Sigma$ and
the pion decay constant $F_\pi$. 
Hence the partition function in this limit is {\sl universal} in
the sense that all theories with a given symmetry breaking pattern
and a mass gap
will have a partition
function given by the same static integral over the 
Goldstone manifold provided that
$m_\pi,\,\mu \ll 1/L \ll \Lambda$ \cite{Vplb,OTV,DOTV,AD4}. 
The simplest theories in this class are invariant random matrix theories in the
limit of large matrices. Because of the large invariance group it
is sometimes simpler to analyze the random matrix theory rather than
to evaluate the integral over the Goldstone manifold directly.

%\vspace{3mm}

In order to derive the microscopic eigenvalue density 
of the Dirac operator, which is defined by rescaling its eigenvalues as
$z_kV\Sigma$,
we start from the microscopic partition function with $N_f+2n$ flavors. Then
we take derivatives with respect to the mass of the additional $n$ quarks and
$n$ conjugate quarks, and finally we remove the additional flavors by taking 
the limit $n\to0$. This procedure is known as the replica trick \cite{EA}.  
Since the microscopic partition functions are only known for integer $n$
there is no guarantee that a correct nonperturbative answer is obtained
this way. After a two decade long discussion \cite{replica} two closely
related methods have emerged that result in the correct nonperturbative 
results: the replica limit of the Painlev\'e equation 
\cite{kanzieper02,kanzieper03} and the replica limit of the 
Toda lattice equation \cite{SplitVerb1,SplitVerb2}.
Both the Painlev\'e equation and the Toda lattice equation are well-known
in the theory of exactly solvable systems (see for example 
\cite{Korepin,ForresterBook}). The Painlev\'e equation is a complicated
nonlinear differential equation, whereas the Toda lattice equation
is a simple two step recursion relation. For that reason, it is much 
simpler to work with the Toda lattice equation. The advantage of the 
Painlev\'e equation is that nonperturbative results can be obtained from
fermionic partition functions only \cite{kanzieper02}.
Recently,
 the consistency of the replica limit of the Toda lattice
equation \cite{SplitVerb3} 
and the supersymmetric method \cite{Efetov} was established.
A necessary ingredient for the applicability of the Toda lattice method
is that the QCD partition function satisfies the Toda lattice equation.
This is the case if the QCD partition function is a $\tau$-function
\cite{ForresterBook}. For the low-energy limit of QCD
at $\mu=0$ this was shown for fermionic partition functions in 
\cite{Kharchev,ADII}. These results were extended to 
bosonic and supersymmetric partition functions as well
as to QCD at nonzero chemical   
potential in \cite{SplitVerb1,SplitVerb2,SplitVerb3}.

In the replica limit of the Toda lattice equation, generating functions 
with bosonic quarks appear in addition to generating functions with fermionic
quarks. While the fermionic generating functions for
the spectral density at nonzero baryon chemical potential 
are known from the integral over the 
Goldstone manifold \cite{SplitVerb2,AFV}, only the simplest case with bosonic
quarks is known \cite{SplitVerb2}. The relevant supersymmetric
generating functions, 
for which the Goldstone manifold is a supermanifold,
 have proved to be quite challenging to compute, and no
explicit expressions have been obtained so far. 
In the present paper we derive these generating  
functions from the random matrix model introduced in \cite{O}. The
flavor symmetries 
and their spontaneous and explicit breaking are exactly the same
in the large $N$ limit of this random matrix model and in
full QCD. Since both theories have a mass gap, in the microscopic
limit, they are both given by the same integral over the Goldstone
manifold. 

In the framework of random matrix models, spectral correlation
functions can also be computed by means of (bi-)orthogonal 
polynomials in the complex plane. The complex orthogonal 
polynomial approach was developed in 
\cite{francesco,FKS,gernotSpectra,AV,BII,AP} and was applied to 
unquenched \cite{O} QCD Dirac spectra at nonzero  chemical potential. 
The outcome of our analysis here is that both the Toda lattice and the 
orthogonal polynomials approach produce equivalent results. 
The divergence of the 
bosonic generating  functions
also sheds some light on universality of non-Hermitian random
matrix models in general.

%\vspace{3mm}

Spectra of the non-Hermitian Dirac operator at nonzero baryon density have been obtained from
lattice QCD in the quenched case \cite{Barbour,tilomar,hands,GT} and for QCD
with two colors \cite{bittner,gbittner} and have been compared successfully
to random matrix theory \cite{tilomar,hands,GT,gbittner}. The microscopic
spectral density of quenched lattice QCD at nonzero baryon density was first
analyzed in \cite{GT} where quantitative agreement with 
analytical predictions \cite{gernotSpectra,SplitVerb2} was found in an
asymptotic domain where 
the results derived from the chiral Lagrangian \cite{SplitVerb2}
agree with the expression obtained in \cite{gernotSpectra}.
The two flavor phase quenched partition function which is calculated
in this paper 
does not suffer from a sign problem either and could be simulated
on a lattice (see \cite{Kogut-Sinclair,Nakamura} for recent results). 
A first analytical  prediction 
for the unquenched QCD Dirac spectrum at nonzero baryon
chemical potential in the non-perturbative regime was obtained in \cite{O}. 
These results and the work presented in this paper 
could in principle be 
compared to numerical simulations in the region where the sign problem sets in.
In particular, any proposal for a solution of the sign problem on the lattice 
can be tested against these predictions.

For the spectral 
density with one dynamical flavor we show both analytically and 
numerically that large fluctuations  occur for increasing  
value of the chemical potential. These findings further illustrate the 
difficulties encountered when one wishes to reproduce these results 
in lattice  QCD simulations.

%\vspace{3mm}

In section \ref{sec:densdef} we explain the derivation of the spectral density 
of the QCD Dirac operator within the replica framework. The required
generating functions are all given in terms of group integrals 
that represent low-energy effective QCD-like partition functions.
The bulk of this paper is devoted to the  calculation of  the group integrals
corresponding to bosonic quarks using the large $N$ limit of a random
matrix model.
After introducing the relevant random 
matrix model for QCD with chemical potential
in section \ref{QCDMM}, we compute the necessary generating functions by
means of complex orthogonal polynomials
in section \ref{MMgenerating}.  
In section \ref{todaop} we show that these partition 
functions satisfy the Toda lattice 
equation. This allows us to calculate the spectral density from
the replica limit of the Toda lattice equation. We show in general terms
that the results agree with a direct
computation using orthogonal polynomials \cite{O}. 
In section \ref{results} we present our results for one, two and any number 
of dynamical flavors and compare them to the quenched and phase quenched 
results. It also contains a discussion of the thermodynamic
limit of our results. This section is self contained and can be read
independently of the previous sections.
Our main findings are
summarized in the conclusions, and several 
technical details are worked out in two appendices.

%%%%%%%%%%%%%%%%%%%%%%%%%%%%%%%%%%%%%%%%%%%%%%%%%%%%%%%%%%%%%%%%%%%%%%%%%%%%%%
\section{The spectral density from generating functions}
\label{sec:densdef}

At nonzero baryon chemical potential the Dirac operator is non-Hermitian so 
that its 
spectrum has a two-dimensional support in the complex plane.
% (provided we choose a suitable regularization such as the lattice). 
In this section we start with the definition of 
the spectral density for complex eigenvalues. Then we 
show how it can be obtained from a generating function
with additional flavors.

%\vspace{4mm}

We consider the eigenvalues of the Dirac operator given by the eigenvalue
equation 
\be
(D_\eta\gamma_\eta + \mu\gamma_0)\psi_j=  z_j\psi_j. 
\label{Dirac}
\ee 
Since 
$(D_\eta\gamma_\eta)^\dagger=-D_\eta\gamma_\eta$ and
$(\mu\gamma_0)^\dagger=\mu\gamma_0$ the Dirac operator is neither 
Hermitian nor anti-Hermitian and the eigenvalues are complex.
Because of the axial symmetry
$\{D_\eta\gamma_\eta + \mu\gamma_0,\gamma_5\}=0$
the nonzero eigenvalues come in pairs with opposite sign, $\pm z_j$.
The density of eigenvalues in the presence of $N_f$ flavors 
is defined as the vacuum expectation value 
over a sum of delta functions in the complex plane  
at the positions of the eigenvalues, vanishing and non-vanishing, 
\be
\label{defdens}
\rho_\nu^{N_f}(z,z^*,\{m_f\};\mu)+\nu\delta^{(2)}(z)
\equiv \left\langle \sum_{j}\delta^2(z-z_j)\right \rangle_{\rm QCD,\,\nu},
\ee
where $\mu$ is the chemical potential, $N_f$ is the number of flavors and
$\{m_f\}=m_1,\ldots,m_{N_f}$ are the quark masses. For later
convenience we have not included the zero 
eigenvalues in our definition of the spectral density. 
The average over gauge fields in a fixed topological sector, $\nu$, is
defined by
\be
\left\langle{\cal O}\right\rangle_{{\rm QCD},\, \nu}\equiv\frac{\int[{\rm
    d}A]_\nu\ {\cal O}\ \prod_{f=1}^{N_f}\det(D_\eta\gamma_\eta +
  \mu\gamma_0+m_f) \ \mbox{e}^{-S_{\rm YM}(A)}}{\int[{\rm
    d}A]_\nu \prod_{f=1}^{N_f}\det(D_\eta\gamma_\eta +
  \mu\gamma_0+m_f) \ \mbox{e}^{-S_{\rm YM}(A)}} .
\ee 
Since the phase of the product of the fermion determinants is non-vanishing,
such expectation values of operators will in general be complex. 
In particular we will find that 
the unquenched spectral density is complex rather than real non-negative. 

The two-dimensional 
$\delta$-functions appearing in the definition of the eigenvalue density
can be obtained from the derivative
\be
\label{densfromgen}
\left\langle \sum_{j}\delta^2(z-z_j)\right \rangle_{\rm QCD,\,\nu} 
& = & \frac{1}{\pi}
\del_{z*} G^{N_f}_\nu (z,z^*,\{m_f\};\mu),
\ee
with the resolvent defined by the average
\be
G^{N_f}_\nu (z,z^*,\{m_f\};\mu) =
\left\langle\sum_j \frac 1{z+z_j}
\right\rangle_{{\rm QCD}, \nu}.
\ee
One way to calculate the resolvent is by means of the fermionic replica trick.
In the usual replica trick the resolvent is calculated from the
identity  
\be
\sum_j\frac 1{z+z_j} = \lim_{n\to 0}\frac 1n \del_z \prod_j(z+z_j)^n\,.
\ee
The problem is that, in most cases, the average of the right hand side can only
be obtained for nonzero integer values of $n$, and the limit $n \to 0$ can
only be taken after a proper analytical continuation in $n$. 
For a non-Hermitian
operator this procedure does not work. Instead, a modified
identity has been suggested \cite{Girko}
 \be
\sum_j\frac 1{z+z_j} = \lim_{n\to 0}\frac 1{n} \del_z \prod_j|(z+z_j)|^{2n}.
\ee 
At the perturbative level the correct resolvent can be derived this way
\cite{Girko,misha}. However, for nonperturbative calculations this relation
only leads to correct results if it is used in combination with the 
Toda lattice equation \cite{SplitVerb1}.

The generating function for the resolvent is thus given by
\be
\label{repgenQCD}
{\cal Z}^{N_f,n}_\nu(\{m_f\},z,z^*;\mu) = \int[{\rm
    d}A]_\nu \prod_{f=1}^{N_f}
\det(D_\eta\gamma_\eta + \mu\gamma_0 + m_f)\ |\det(D_\eta\gamma_\eta +
\mu\gamma_0 + z)|^{2n} \ \mbox{e}^{-S_{\rm YM}(A)}.
\ee
The resolvent is obtained by
\be
G^{N_f}_\nu (z,z^*,\{m_f\}; \mu) = \lim_{n \to 0} \frac 1{n} 
\del_z \log{\cal  Z}^{N_f,n}_\nu(\{m_f\},z,z^*;\mu),
\ee
and the spectral density is given by
\be
\rho^{N_f}_\nu(z,z^*,\{m_f\};\mu)+\nu\delta^{(2)}(z) &=& 
\lim_{n\to0} \frac{1}{\pi n}\partial_{z^*}\partial_z
\log{\cal  Z}^{N_f,n}_\nu(\{m_f\},z,z^*;\mu). 
\label{replica}
\ee
This construction has been used previously for perturbative \cite{Girko,misha}
and nonperturbative \cite{SplitVerb2,kanzieper03} studies of the eigenvalue
density in quenched theories (i.e. for $N_f=0$). 

The interpretation of the partition function (\ref{repgenQCD}) 
is \cite{Gocksch,misha}
that in addition to the usual quarks we have
$n$ quarks with mass $z$ and $n$ conjugate quarks with mass $z^*$.
Since
\be\label{conj-iso}
  \det(D_\eta\gamma_\eta+\mu\gamma_0+z)^* 
= \det(\gamma_5(-D_\eta\gamma_\eta+\mu\gamma_0+z^*)\gamma_5)
= \det(D_\eta\gamma_\eta-\mu\gamma_0+z^*),
\ee
conjugate quarks carry  the opposite baryon number of quarks.
The baryon charge matrix in the partition function (\ref{repgenQCD})
is therefore not proportional to the identity and the flavor symmetry 
is broken by the chemical potential.
The specific way in which the flavor symmetries are broken determines the
low-energy effective theory. In particular, the
effective theory will depend on $\mu$ for energies where the pions
dominate the free energy. Physically this is clear because mesons made
out of quarks and conjugate anti-quarks carry a nonzero baryon number. 
The general prescription
for constructing a chiral Lagrangian is to impose the 
transformation properties of the microscopic theory on the low-energy
effective theory \cite{GL}. For an external vector field the microscopic
partition function is invariant under local gauge transformations of
this field. Therefore, the derivative in the chiral Lagrangian has
to be replaced by the corresponding covariant derivative \cite{GL,KST}. 

In this paper we calculate the eigenvalue density 
in the microscopic limit for any number of flavors $N_f$, 
where $m_f V\Sigma$, $z V\Sigma$, and $\mu^2F_\pi^2
V$ are held fixed as $V\to\infty$. In this limit the kinetic terms
factorize from the partition function and the
low energy limit
of the generating function (\ref{repgenQCD}) is given by \cite{dominique-JV}
\be
Z^{N_f,n}_\nu(\{m_f\},z,z^*;\mu) = 
\int_{U \in U(N_f+2n)} dU \ \det(U)^\nu\ \mbox{e}^{-\frac {V}{4}F_\pi^2\mu^2
{\rm Tr} [U,B][U^{-1},B]\ +\ \frac 12 \Sigma V {\rm Tr}M(U + U^{-1})}.
\label{zeff}
\ee
The quark mass matrix is given by
$M$=diag$(m_1,\ldots,m_{N_f},\{z\}_n,\{z^*\}_n)$  
and the charge matrix $B$ is a diagonal 
matrix with 1
appearing $N_f+n$ times along the diagonal and $-1$ appearing $n$ times.
An explicit expression for this integral was derived in \cite{SplitVerb2,AFV}.
 In this paper
we will show that the partition functions (\ref{zeff}) satisfy the Toda lattice
equation
\be 
\frac 1{\pi n} \delta_z\delta_{z^*} \log Z^{N_f,n}_\nu(\{m_f\},z,z^*;\mu) 
= \frac{1 }2 
(zz^*)^2 
\frac{Z^{N_f,n+1}_\nu(\{m_f\},z,z^*;\mu)Z^{N_f,n-1}_\nu(\{m_f\},z,z^*;\mu)}
{[Z^{N_f,n}_\nu(\{m_f\},z,z^*;\mu)]^2} ,
\label{TodaBaryon}
\ee
where $\delta_z = z d/dz$ and $\delta_{z^*} = z^*d/dz^*$.
Since the replica limit of the right hand side of this equation is the 
spectral density\footnote{Notice that both in the l.h.s. and the
r.h.s. the overall factors $z$ and $z^*$ are cancelled so that the
l.h.s. of (15) gives the spectral density without the zero
eigenvalues.} we find the
remarkably simple expression  \cite{SplitVerb1,SplitVerb2,SplitVerb3} 
\be
\rho^{N_f}_\nu(z,z^*,\{m_f\};\mu) &=& \frac {zz^*}2
\frac{Z^{N_f,n=1}_\nu(\{m_f\},z,z^*;\mu)Z^{N_f,n=-1}_\nu(\{m_f\}|z,z^*;\mu)}
{[Z^{N_f, n=0}_\nu(\{m_f\};\mu)]^2}. 
\label{rhoToda}
\ee
In (\ref{rhoToda}) and elsewhere in this paper
it is our notation to separate the fermionic and bosonic
quark masses in the argument of the partition function by a vertical
line.
In addition to the fermionic partition functions, 
we will need to evaluate $Z_\nu^{N_f,n}$  for
$n=-1$, i.e. for one  bosonic quark and one bosonic conjugate quark. 
In the quenched case ($N_f=0$), this partition function
is given by the integral \cite{SplitVerb2,minn04}
\be
Z_\nu^{n=-1}(z,z^*;\mu) = 
\lim_{\epsilon\to0} C_\epsilon
\int \frac{dU}{{\det}^{2-|\nu|} U} \theta(U) \ 
\mbox{e}^{ -\frac{V}{4}F_\pi^2 \mu^2 {\rm Tr} [U ,B][U^{-1} ,B ]
\ +\ \frac{i}{2} {V\Sigma}{\rm Tr} M^T(U -I U^{-1}I )   } ,
\label{ZMINQCD}
\ee 
where $dU\theta(U)/{\det}^2 U $ is the integration measure on positive
definite $2\times 2$ Hermitian matrices and
\be
B = \mat 1 & 0 \\ 0 & -1 \emat \qquad , \qquad M = \mat \epsilon & z
\\ z^* & \epsilon \emat  \qquad {\rm and} 
\qquad I = \mat 0 & 1 \\ -1 & 0 \emat.
\ee
It was found that in order to obtain a finite limit $\epsilon \to 0$,
the normalization constant has to be chosen $C_\epsilon \sim 1/\log\epsilon$.
The appearance of this logarithmically diverging 
term is not surprising since the
inverse fermion determinants are regularized as
\be
{\det}^{-1}(D_\eta \gamma_\eta +\mu\gamma_0 +z)
{\det}^{-1}(-D_\eta\gamma_\eta+\mu\gamma_0 +z^*)
= \lim_{\epsilon \to 0} {\det}^{-1}\mat \epsilon &  D_\eta\gamma_\eta
+\mu\gamma_0 +z\\
            D^\dagger_\eta\gamma_\eta+\mu\gamma_0 +z^* & \epsilon \emat.
\label{regdet}
\ee
Zero eigenvalues of $D_\eta \gamma_\eta$
do not lead to  zero eigenvalues of the operator in
the r.h.s. of this equation. Therefore, generically its eigenvalues
$\lambda_k$ are nonzero real and occur in pairs $\epsilon \pm \lambda_k$, and
we expect that the $\epsilon$ dependence enters in a similar way to the mass
dependence of the QCD partition function with
one bosonic quark at zero topological charge. In the microscopic limit this 
is given by $K_0(m)$ which diverges logarithmically in $m$. 

To derive the partition functions (\ref{zeff}) and (\ref{ZMINQCD})
we have only used the global symmetries of the underlying QCD partition 
functions.  In particular, this implies that any microscopic theory
with the same global symmetries and spontaneous breaking thereof will
have the same zero momentum effective theory. The simplest
theory in this class is chiral Random Matrix Theory at nonzero chemical
potential which is obtained from the QCD partition function by replacing
the matrix elements Dirac operator by (Gaussian) random numbers.
The partition function (\ref{ZMINQCD}) was also
derived explicitly starting from a chiral random matrix model instead of
only using symmetry arguments \cite{SplitVerb2}.
Another advantage of using a random matrix model is that one can 
easily perform numerical simulations. For example, the quenched spectral
density was calculated numerically \cite{SplitVerb2} and was
found to be in agreement with (\ref{rhoToda}). 

Below we will show that the replica limit of the Toda lattice equation
(\ref{rhoToda}) can also be used to obtain the
unquenched spectral density using a similar regularization of the partition
function with bosonic quarks. In this case we need
$Z_\nu^{N_f,n=-1}$ with $N_f\neq0$. 
This involves an integral over the supergroup
$\hat{Gl}(N_f|2)$. This integral is not known explicitly for $N_f\neq0$.
For that reason we will derive $Z_\nu^{N_f,n=-1}$ directly from the
corresponding 
random matrix model instead of  the low-energy effective partition function
based on $\hat{Gl}(N_f|2)$.

%%%%%%%%%%%%%%%%%%%%%%%%%%%%%%%%%%%%%%%%%%%%%%%%%%%%%%%%%%%%%%%%%%%%%%%%%%%%%%
\section{Random matrix model for QCD at nonzero chemical potential}
\label{QCDMM}

 In this section we define the random matrix model which we will use
to calculate the generating  functions. Random matrix models for
QCD originally \cite{SV} focused on 
the quark mass dependence at zero chemical potential. 
Later \cite{misha}, a random matrix model including
the chemical potential  successfully explained 
why quenched lattice QCD at zero temperature
has a
phase transition at a chemical potential of half the pion mass. 
However, a disadvantage of this model is that no 
eigenvalue representation is known which is required for the use
of orthogonal polynomial methods.
The random matrix model that will be used in the present
paper was introduced in \cite{O}. It differs from the model in \cite{misha}
by a different form of the 
chemical potential term. Its partition function can be reduced to an 
eigenvalue representation and  
allows for an explicit solution in terms of orthogonal polynomials \cite{O}.
Because this model captures the correct 
global symmetries of the QCD Dirac operator, its low energy limit is also given
by (\ref{zeff}). 
A related model defined in terms of a joint eigenvalue distribution
was introduced in \cite{gernotSpectra}. 
Although this model is not in the universality class of QCD partition
 functions,
the complex orthogonal polynomial methods that were developed 
for the derivation
of the spectral density are applicable to the 
random matrix model introduced in \cite{O}. 
In the section  \ref{todaop} we
will comment on  the relations between  the different models. 

%\subsection{Definition of the Random Matrix Model}

The random matrix partition function with $N_f$ quark flavors of mass $m_f$
and $n$ replica pairs of regular and conjugate quarks with masses $y$ and $z^*$
each is defined by  
\be
{\cal Z}_N^{N_f,n}(\{m_f\},y,z^*;\mu) &\equiv&
 \int d\Phi  d\Psi \ w_G(\Phi)  w_G(\Psi) 
\prod_{f=1}^{N_f} \det(\,{\cal D}(\mu) + m_f\,) \  \nn \\
&&\times
{\det}^n(\,{\cal D}(\mu) + y\,){\det}^n(\,{\cal D}^\dagger(\mu) + z^* ) ,
\label{ZNfNb}
\ee
where the non-Hermitian Dirac operator is given by
\be
\label{dnew}
\mathcal{D}(\mu) = \left( \begin{array}{cc}
0 & i \Phi + \mu \Psi \\
i \Phi^{\dagger} + \mu \Psi^{\dagger} & 0
\end{array} \right) ~.
\ee
Negative numbers of flavors $N_f,n<0$ will denote insertions of 
bosons given by inverse powers of determinants. 
To allow for the presence of both bosonic and fermionic flavors
we will distinguish between $N_b$ and $N_f$ in later formulas.
Here $\Phi$ and $\Psi$ are complex $(N+\nu)\times N$ matrices with the same
Gaussian weight function
\be
\label{wg}
w_G(X) ~=~ \exp( \, - \, N \, \tr \, X^{\dagger} X \, ) ~.
\ee 
The matrix $\mathcal{D}(\mu)$ replaces the QCD Dirac operator plus chemical 
potential in (\ref{Dirac}). It has exactly $\nu$ zero modes which 
identifies $\nu$  as the absolute value of the topological charge. 
The model (\ref{ZNfNb}) has the same
flavor symmetries as QCD which are broken by
the quark masses and the chemical potential in
exactly the same way as in QCD. 
The only difference with the model
\cite{misha} is that
the chemical potential is multiplied by a  complex matrix
$\Psi$ with Gaussian distributed matrix elements. 
In the
microscopic limit, where $\mu^2N$, $m_f N$ and $z N$ are fixed as
$N\to\infty$ this partition function will match  the
partition function (\ref{zeff}) after fixing the
scale of the parameters appropriately. The latter is uniquely determined 
in terms of the flavor symmetries and their breaking, and thus universal. 
As was already done in the previous
section, the microscopic limit of the partition function will be
denoted by $Z$ instead of the notation ${\cal Z}$ which we use
both for the QCD partition function and the random matrix theory
partition function. Generally, the overall normalization
of $Z$ and the microscopic limit of ${\cal Z}$ is different. 

The form of the Gaussian weight
$w_G(\Phi)$ respects the flavor symmetries, but these symmetries do not
exclude traces of higher powers of $\Phi^{\dagger}\Phi$ in the exponent
as well as other non-invariant terms. 
At zero
chemical potential it was shown 
\cite{JSV-1,ADMN,Guhr-tilo,JSV-2,DN,FK,DV,GF} that 
partition functions and eigenvalue correlations in the microscopic limit are
independent of the form of this weight, which, in random matrix theory,
is known as universality. The only condition for the probability density is
that the spectral density near the origin is nonvanishing in the 
thermodynamic limit.
In the derivation of the joint eigenvalue density of (\ref{ZNfNb})
it is essential that the two weight functions are Gaussian.
However, we believe that this is only a technical requirement and higher
order invariant terms will not alter the microscopic limit of the
joint probability distribution
(for more discussion see section \ref{todaop}).

%\subsection{Eigenvalue representation}

In \cite{O} it was shown that the partition function (\ref{ZNfNb}) 
has an eigenvalue representation after choosing an appropriate representation
for the matrices $\Phi$ and $\Psi$ yielding
\be
\label{epfnew}
{\cal Z}_N^{N_f,n}(\{m_f\},y,z^*;\mu)  \sim
(yz^*)^{\nu n} \prod_{f=1}^{N_f} m_f^{\nu}\ 
 \int_{\mathbf{C}} \prod_{k=1}^{N} d^2z_k \,
{\cal P}^{N_f,n}_\nu(\{z_i\},\{z_i^*\}; a),
\ee
where the  integration extends over the full complex plane and 
the joint probability distribution of the eigenvalues is given by
\be
\label{jpd}
{\cal P}^{N_f,n}_\nu(\{z_i\},\{z_i^*\};a)
&=& \frac{1}{\mu^{2N}}\left|\Delta_N(\{z_l^2\})\right|^2 \, 
\prod_{k=1}^{N} w^{N_f,n}_\nu(z_k,z_k^*;{a}).
\ee 
The Vandermonde determinant is defined as  
\be
\Delta_N(\{z^2_l\}) \equiv \prod_{i>j=1}^N (z_i^2-z_j^2),
\label{vander}
\ee
and the weight function reads
\be 
w^{N_f,n}_\nu(z_k,z^*_k;{a}) &=&
[\prod_{f=1}^{N_f} \, (m_f^2-z^2_k )]\, (y^2-z_k^2 )^{n}
(z^{*\,2}-z_k^{*\,2} )^{n}  
|z_k|^{2\nu+2} 
K_\nu \left( \frac{N (1+\mu^2)}{2 \mu^2} |z_k|^2 \right)
\nn \\ &&\times \exp\left(-\frac{N (1-\mu^2)}{4 \mu^2}  
(z^2_k + \conj{z_k}^2) \right), 
\label{wnew}
\ee
where $K_\nu$ is a modified Bessel function. The parameters of the
weight function, $\{m_f\}, y,z^*$ and $\mu$ are collectively denoted 
by ${a}$.
In the quenched case, $N_f=n =0$, the weight function will be denoted
by $w(z_k,z^*_k;\mu)$. We have included a normalization factor 
$1/\mu^{2N}$ in eq. (\ref{jpd}). This way it reduces to the joint 
eigenvalue density  of the chGUE times a product of delta-functions 
in the limit $\mu\to 0$. We expect that the
terms that are nonvanishing in the microscopic limit are universal
and are required for the derivation of the microscopic spectral density.

In the asymptotic limit, $N|z_k|^2/\mu^2 \gg 1$, the modified Bessel function
in  
the weight is well approximated by 
its leading order asymptotic expansion. The corresponding 
joint eigenvalue distribution was introduced in \cite{gernotSpectra}
as a model for QCD at nonzero chemical potential. Let us look closer
at the condition $N|z_k|^2/\mu^2 \gg 1$.
In terms of the half-width, $x_{\rm max}\sim\mu^2$, 
of the cloud of eigenvalues and
the spacing, $\Delta \sim 1/N$, of the imaginary part of the eigenvalues it can
be rewritten as
\be
|z_k|^2 / (x_{\rm max }\Delta) \gg 1.
\ee
For $x_{\rm max} > \Delta$  this implies that $|z_k| \gg \Delta$ where we
expect to find the weight function of the
Gaussian Unitary Ensemble perturbed by an anti-Hermitian matrix which indeed
has a Gaussian weight function (see \cite{FKS}). In other words, the
appearance of the modified Bessel function $K_\nu$ is a direct consequence
of the chiral symmetry of the problem.

%%%%%%%%%%%%%%%%%%%%%%%%%%%%%%%%%%%%%%%%%%%%%%%%%%%%%%%%%%%%%%%%%%%%%%%%%%%%
\section{Generating functions from random matrix models}
\label{MMgenerating}

The aim of this section is to derive explicit expressions for the 
generating functions introduced in section 
\ref{sec:densdef} using the random matrix 
model introduced in the previous section.
In particular, we will present new expressions for 
the corresponding partition
functions with any given number of fermions and one pair of conjugate bosonic
quarks, as they are needed for the replica limit of the Toda lattice equation.
We will find that the partition functions with bosonic replicas 
have to be regularized. This
divergence will lead to a factorization 
into a partition function containing only fermions 
and a purely bosonic one. 
After taking the microscopic  limit, 
%\ref{sec:micro}
they can
be expressed entirely in terms of  known partition functions 
(\ref{zeff}) and (\ref{ZMINQCD}).
In the following 
section we consider  partition functions with  $N_f$ flavors and
$n$ fermionic replicas to show that the random
 matrix model partition functions 
satisfy the Toda lattice equation (\ref{TodaBaryon}).  

\subsection{Orthogonal polynomials and Cauchy transforms}

%\subsection{Partition functions with one flavor}

As is the case for Hermitian random matrix models,
the partition functions (\ref{epfnew}) can be expressed in terms 
of orthogonal polynomials, and the integrals to obtain the spectral
correlation function
can be performed by means of the orthogonality relations. In this 
section we present the polynomials corresponding to the 
weight function (\ref{wnew}) for $N_f = n =0$ (which is denoted
by $w(z,z^*;\mu)$). 

The orthogonal polynomials are defined as solutions to the following
orthogonality relation 
\be
\mathop{\int}_{\mathbf{C}}d^2z\ w(z,z^*;\mu)\ p_k(z)\ p_l(z)^* ~ 
 \ =\ r_k ~ \delta_{kl} \ ,
\label{OPdef}
\ee
where the integral is over the full complex plane with measure
$d^2 z = d{\rm Re}z\, d{\rm Im} z$,
and $r_k$ denotes the 
squared norms of the polynomials. Since the quenched weight 
$w(z,z^*;\mu)$
appearing in this relation is real positive the polynomials form a complete 
set with positive squared norms $r_k$. The polynomials $p_k(z)$ depend 
only on the variable $z$ and not its complex conjugate, as is indicated by 
the notation. They are also functions of
the chemical potential $\mu$, the 
size $N$ of the random matrix, as well as of the topological charge $\nu$. We
suppress the dependence on $\mu$, $N$ and $\nu$.

In monic normalization, 
$p_k(z)=z^{2k}+\ldots$, the solution of (\ref{OPdef}) is given
by \cite{O} 
\be
\label{Laguerre}
p_k (z) = 
\left( \frac{1-\mu^2}{ N} \right)^{k} k! ~
L_k^{(\nu)}\left(-\frac{ N z^2}{1-\mu^2}\right).  
\ee
Since the Vandermonde determinant only contains squared variables 
only  polynomials in those squared variables will appear. 
The same happens for the polynomials on the real line when mapping the 
Laguerre ensemble from the positive real numbers to the full real line. 
For the weight (\ref{wnew}) the norms are given by
\be
\label{Norm}
r_k ~=~
\frac{  \pi \, \mu^2 ~ (1+\mu^2)^{2k+\nu} ~ k! ~ (k+\nu)!}
     { N^{2k + \nu + 2}}  ~.
\ee
Because the Laguerre polynomials $L_k^{(\nu)}(z)$ 
have real coefficients it holds that
$p_k(z)^*=p_k(z^*)$. 

The Cauchy transform of the polynomial $p_k(z)$ is defined by 
\be
\label{Cauchydef}
h_k(y) \equiv  \int_{\mathbf{C}} d^2z\ w(z,z^*;\mu)\ p_{k}(z)^* \ 
\frac{1}{z^2-y^2}.  
\ee
We note that $h_0(y)$ has a nontrivial $y$-dependence, 
unlike the polynomial $p_0(y)$.
In contrast to random
matrix models with eigenvalues on a curve in the complex plane,
the integral over the pole in the complex plane is always well defined. 
However, if we take the limit $\mu\to 0$ of an anti-Hermitian Dirac operator
we have to make sure that the argument $y$ lies outside the support of 
eigenvalues.

The leading order asymptotic large-$y$ behavior of $h_k(y)$ is obtained
by expanding $1/(y^2-z^2)$ in a geometric series. Because of
orthogonality all coefficients up to order $1/y^{2k}$ vanish resulting
in the asymptotic behavior
\be
h_k(y) = -\frac{r_k}{y^{2k+2}} + O(\frac 1{y^{2k+4}}).
\label{asymcauchy}
\ee
This result is useful for checking the identities  
derived below.

\subsection{Partition functions with one flavor}

The polynomials and their Cauchy transform enjoy the following relations 
to one flavor partition functions. The former is related to a single 
fermionic flavor partition function through the so-called Heine formula,
\be
\frac{{\cal Z}_N^{N_f=1}(z;\mu)}{{\cal Z}_N^{N_f=0}} \equiv 
\left\langle\ z^\nu \prod_{j=1}^N (z^2-z_j^2 )\ \right\rangle_{N_f=n=0}
= z^\nu p_N(z) \ . 
\label{Heine}
\ee
Here we have written the fermion determinant as an expectation value
with respect to the quenched partition function at $N_f=n=0$.
(Vanishing replica (or flavor) indices will be commonly suppressed throughout
the following.) For $ \mu < 1$ 
the partition function is manifestly real and positive 
for real masses $z$, as the 
orthogonal polynomials are Laguerre polynomials of argument
$-{ N z^2}/{(1-\mu^2)}$ with zeros on the imaginary axis.
For $\mu>1$ the zeros are located on the real axis. Remarkably, the 
Yang-Lee zeros \cite{HJV} of the model in \cite{misha} behave
quite different from the model (\ref{ZNfNb}) discussed in this paper.
The polynomials can also be interpreted as characteristic polynomials.

The relation (\ref{Heine}) can also be written down for 
$p_{l\neq N}(z)$, expressing it as 
an average over $l$ variables. However, in order to be precise 
we would have to 
chose the weight function in (\ref{wnew})  to be 
$N$-independent as in \cite{AV,BII,AP}. Since we are mainly interested 
in taking the large-$N$ limit we prefer to explicitly keep the $N$ or 
volume dependence inside the weight. 
For $l\approx N$ we can safely ignore this technical subtlety.

Turning to the Cauchy transform of $p_N(z)$, it is 
given by the partition function with one bosonic quark
flavor \cite{BII,AP}, 
\be
\frac{{\cal Z}^{N_f=-1}_N(z;\mu)}{{\cal
    Z}_N^{N_f=0}} \equiv 
\left\langle\ z^{-\nu}\prod_{j=1}^N \frac{1}{(z^2 -z_j^2 )}\ \right
\rangle_{N_f=n=0} =\ 
 \frac{-1}{r_{N-1}}z^{-\nu} \ h_{N-1}(z)\  .
\label{CHeine}
\ee
The same remarks about positivity and the $N$-dependence of the weight function
made before also apply here.

Our program is to express spectral correlation functions in terms of 
partition functions. This has the advantage that the microscopic limit
is given by a group integral based on the symmetries of the partition
functions. To achieve our goal we also need all possible two flavor
partition functions. At zero chemical potential the program of expressing
spectral correlation functions in terms of  multiflavor partition functions
was carried out in  \cite{AD,Fyodor-ch}.
%%%%%%%%%%%%%%%%%%%%%%%%%%%%%%%%%%%%%%%%%%%%%%%%%%%%%%%%%%%%%%%%%%%%%%

\subsection{Partition function with two flavors}

It is known from Hermitian random matrix models \cite{PZJ} 
that the expectation value 
of two characteristic polynomials is proportional
to the kernel of the orthogonal polynomials (\ref{OPdef}). 
As was shown in \cite{AV}, 
for an arbitrary weight function, similar results can be derived for
non-Hermitian random matrix models.
Since we are
now dealing with non-Hermitian matrices we have to distinguish between 
a characteristic polynomial (or mass term) and its Hermitian conjugate. 
For a single pair of 
one fermionic quark with mass $z$ and one conjugate
fermionic quark (with complex conjugated eigenvalues)
with mass $u^*$ we have the relation \cite{AV}
\be
\frac{{\cal Z}_N^{n=1}(z,u^*;\mu)}{{\cal Z}_N^{N_f=0}} &\equiv& 
\left\langle\ (zu^*)^\nu\prod_{j=1}^N (z^2 -z_j^2 )(u^{*\, 2}-z_j^{*\,2})
\ \right\rangle_{N_f=n=0} \ =\  r_{N}\ (zu^*)^\nu K_{N+1}(z,u^*),\ 
\label{KpreZ}\\
K_{N+1}(z,u^*)&\equiv& \sum_{j=0}^{N}\frac{1}{r_{j}}\, p_{j}(z)\,p_{j}(u^*)\ .
\label{Kdef}
\ee 
Here we have defined the bare kernel $K_N(z,u^*)$ which differs from the
full kernel by the square root of  the weight function at each argument.
The full kernel appears in the computation of 
correlation functions from orthogonal polynomials (see 
 section \ref{todaop}).
In particular, for $z=u$ we obtain the result for one pair of 
fermionic replicas $n=1$
\be
\frac{{\cal Z}_N^{n=1}(z,z^*;\mu)}{{\cal Z}_N^{N_f=0}}\ =\  
\left\langle\ |z|^{2\nu}\prod_{j=1}^N |z^2-z_j^2|^2\ \right\rangle_{N_f=n=0}
\ =\ r_{N}\ |z|^{2\nu}K_{N+1}(z,z^*) \ .
\label{KZ}
\ee
This quantity is manifestly real and positive. 
On the other hand, if we compute the partition function with 
two flavors with the same eigenvalues, we obtain \cite{AV}
\be
 \frac{{\cal Z}_N^{N_f=2}(x_1,x_2;\mu)}{{\cal Z}_N^{N_f=0}} &\equiv& 
\left\langle\ (x_1x_2)^\nu\prod_{j=1}^N (x_1^2-z_j^2 )(x_2^2-z_j^{2})
\ \right\rangle_{N_f=n=0} \ \ =\ \frac{(x_1x_2)^\nu}{(x_2^2-x_1^2)}
\left|
\begin{array}{ll}
p_N(x_1) & p_{N+1}(x_1)\\
p_N(x_2) & p_{N+1}(x_2)\\
\end{array}
\right|\ .
\label{ppZ}
\ee
For Hermitian random matrix models the two results (\ref{KpreZ})  and 
(\ref{ppZ}) are related through the Christoffel-Darboux formula, which 
in general does not hold in the complex plane
(see also \ref{app:Z}).

%\subsection{The partition function with two conjugate bosonic quarks}

We now turn to the most important two-flavor 
partition function, containing   a pair of  a bosonic quark  and its  
conjugate. 
In the replica limit of the Toda lattice equation this 
partition function appears when computing the 
quenched density (see (\ref{rhoToda})).
In the microscopic limit this partition function was calculated in
\cite{SplitVerb2}, both by starting from a random matrix model,  and by
integrating explicitly over the Goldstone manifold in the chiral Lagrangian
(\ref{ZMINQCD}). It turned out that
this partition function is singular and has to be regularized
according to the prescription given in (\ref{regdet}). The partition function
is obtained in the limit of vanishing regulator, $\epsilon \to  0$, where
it diverges as  $\sim \log(\epsilon)$. This diverging constant can be
absorbed in the normalization of the partition function.
Below we will establish
the nature of this singularity in the orthogonal polynomial approach.
In fact, it does not enter in the calculation of the 
spectral density for the random matrix model (\ref{ZNfNb})
by means of complex orthogonal polynomials \cite{O}. However, we will
show that it occurs even at finite $N$ in the random matrix model 
partition function with one bosonic quark and
one conjugate bosonic quark of the same mass.

Generalizing the results in \cite{BII} 
to the chiral case we find that the following relation holds
\be
\frac{{\cal Z}_N^{n=-1}(y,x^*;\mu)}{{\cal Z}_N^{N_f=0}} &\equiv& 
\left\langle\ (yx^*)^{-\nu}
\prod_{j=1}^N \frac{1}{(y^2-z_j^2)( x^{*\,2}-z_j^{*\,2} )}
\ \right\rangle_{N_f=n=0} = \ 
-\frac{(yx^*)^{-\nu}}{r_{N-1}}\ {\cal A}_{N-2}(x^*,y),
\label{AZ}\\
{\cal A}_{N-2}(x^*,y)&\equiv& 
- Q(x^*,y;\mu) + \sum_{j=0}^{N-2} \frac{1}{r_{j}}\,h_{j}(x^*) \, 
h_{j}(y) ,
\label{Adef}
\ee
where 
\be
Q(x^*,y;\mu) &\equiv& \int_{\mathbf{C}} d^2z\ w(z,z^*;\mu) 
\frac{1}{(z^{*\,2}-x^{*\,2})(z^2-y^2)}\ .
\label{Q}
\ee
Equation (\ref{Adef}) defines a bare kernel 
containing only Cauchy transforms. 
For different masses $x\neq y$ both the sum over Cauchy transforms and the
function $Q(x^*,y;\mu)$ are finite.
At equal arguments, $x=y$, the sum remains finite but
the integral in (\ref{Q}) is logarithmically 
divergent, just like the divergence encountered in
(\ref{ZMINQCD}).   
In order to regularize this divergence we cut out a
circle $C(x,\epsilon)$ of radius $\epsilon$ around the pole at $z=x$
and one around the pole at $z=-x$. 
To evaluate the singular part of the 
regularized $Q_\epsilon(x^*,x;\mu)$ we expand the weight function around
the singular point and keep only the leading part. For $|x| \gg \epsilon$ 
we find for the pole at $z = x$
\be
Q_\epsilon(x,x^*;\mu) 
 & \approx &   w(x, x^*;\mu) \int_{\mathbf{C}-C(x,\epsilon)}\frac{dzdz^*}{-2i}
\frac 1{(z^{* 2} -x^{* 2})(z^2 - x^2) } \nn\\
&=&w(x, x^*;\mu) \int_{\mathbf{C}-C(x,\epsilon)}\frac{dzdz^*}{-2i}
\frac 1{2x^*} \del_{z^*}\left [\frac { 
\log((z^*-x^*)/(z^*+x^*))}{z^2-x^2}\right ] \nn\\
& =&w(x, x^*;\mu) \int_{\del C(x,\epsilon)}\frac{dz}{2i}
\frac 1{2x^*}  \frac {
\log((z^*-x^*)/(z^*+x^*))}{z^2-x^2}\nn \\
& =&  
-\frac {\pi w(x, x^*;\mu)}{2 xx^*} \log\epsilon  +O(\epsilon^0).
\label{Qeps}
\ee
We find the same contribution from the pole at $z = -x$.
Thus we have obtained
the following interesting relation
for the singular part of the bosonic partition
function 
\be
\frac{{\cal Z}_N^{n=-1}(x,x^*;\mu)}{{\cal Z}_N^{N_f=0}(\mu)}
\ =\   -\frac{\pi w(x,x^*;\mu)}{2 |x|^{2+2\nu} \ r_{N-1}} \log\epsilon 
+ O(\epsilon^0)\ .
\label{QZ}
\ee
We are not aware that such a relation between 
the weight function (\ref{wnew}) of complex orthogonal polynomials and the 
bosonic partition function at a degenerate mass pair was previously known. 
It translates easily to non-chiral non-Hermitian random matrix models.

Relation (\ref{QZ}) shows that the singularity that was found
in \cite{SplitVerb2} for the microscopic limit of 
${\cal Z}_N^{n=-1}(x,x^*;\mu)$ even persists for finite size random
matrices and can be expressed simply in terms of
the weight function. 
Only weak assumptions have been made in the derivation 
of (\ref{QZ}), and its validity extends to a wide class of weight functions. 
 For $x \sim O(\epsilon)$ the above derivation
does not apply. Instead we find that 
$\lim_{ x \to 0}Q_\epsilon(x,x^*;\mu) \sim 1/\epsilon^2$.
As we will explain in the next section,
the logarithmic singularity we have found for the two-point function 
(\ref{QZ}) will also occur in more general partition functions with
at least one  pair of conjugate bosonic quarks.

The last two-point function we need below
is the partition function with
 one fermion and one boson. This partition function reads
\cite{BII} 
\be
\frac{{\cal Z}_N^{N_f=1,N_b=1}(x|y;\mu)}{{\cal Z}_N^{N_f=0}} &\equiv& 
\left\langle\ \left(\frac{x}{y}\right)^\nu\prod_{j=1}^N \frac{(x^2-z_j^2 )}{(y^2-z_j^{2})}
\ \right\rangle_{N_f=0}=\ (y^2-x^2)\ \left(\frac{x}{y}\right)^\nu {\cal N}_{N-1}(x,y)\ ,
\label{NZ}\\
{\cal N}_{N-1}(x,y) &=& \frac{1}{(y^2-x^2)} +
\sum_{j=0}^{N-1}\frac{1}{r_{j}}\,p_{j}(x)\,h_{j}(y)\ .
\label{Ndef}
\ee
Here, we have defined the  bare kernel ${\cal N}_{N-1}(x,y)$ 
consisting of orthogonal polynomials and 
Cauchy transforms.
This partition function 
is nonsingular for any values of the arguments and correctly normalizes 
to unity at equal arguments $y=x$.

The remaining two flavor partition functions combining 
(conjugate) fermionic quarks and (conjugate) bosonic quarks 
are given in  \ref{app:Z}.

%%%%%%%%%%%%%%%%%%%%%%%%%%%%%%%%%%%%%%%%%%%%%%%%%%%%%%%%%%%%%%%%%%%%%%%%%%%%%

\subsection{Partition Functions with an arbitrary number of flavors
and one conjugate flavor}

General expressions for the expectation value of
ratios of characteristic polynomials valid for an arbitrary complex
weight function were given in \cite{AV,BII,AP}. We will apply these
results
to partition functions with $N_f$ flavors plus 
one pair of regular and conjugate fermionic $(n=+1)$ 
or bosonic $(n=-1)$ flavors which both enter
in the expression for the spectral
density (\ref{rhoToda}) after taking the replica limit $n\to0$.

The  partition function
with $N_f$ fermionic quarks with masses $m_f$ is given by \cite{AV}
\be
\frac{{\cal Z}_N^{N_f}(\{m_f\};\mu)}{{\cal Z}_N^{N_f=0}} \equiv 
\left\langle\  \prod_{f=1}^{N_f}m_f^\nu\prod_{j=1}^N
(m_f^2-z_j^2 )\ \right\rangle_{N_f=n=0}=\ 
\frac{\prod_{f=1}^{N_f}m_f^\nu}{\Delta_{N_f}(\{m_f^2\})}
\det_{1\leq k,l\leq N_f}\left[\ p_{N+k-1}(m_l)\ \right] .
\label{ZNf}
\ee
Next we add a single pair of a fermion and its conjugate. This partition
function is given by \cite{AV}
\be
\frac{{\cal Z}_{N}^{N_f,n=+1}(\{m_f\},z,z^*;\mu)}{{\cal Z}_{N}^{N_f=0}}  
&\equiv&
\left\langle\ |z|^{2\nu}(\prod_{f=1}^{N_f}m_f^\nu)\prod_{j=1}^N \left(|z^2-z_j^2|^2\prod_{f=1}^{N_f}
(m_f^2-z_j^2 )\right)\ \right\rangle_{N_f=n=0} \nn\\
&&\hspace*{-5cm} =
\frac{(-1)^{N_f}r_N|z|^{2\nu}\prod_{f=1}^{N_f}m_f^\nu}{\Delta_{N_f+1}(\{m_f^2\},z^2)} 
\left|
\begin{array}{lll}
p_{N+1}(m_1)     & \ldots\ p_{N+1}(m_{N_f})     & p_{N+1}(z)\\
\hspace{5mm}\vdots & \ddots\ \hspace{5mm}\vdots & \hspace{5mm}\vdots\\
p_{N+N_f}(m_1)   & \ldots\ p_{N+N_f}(m_{N_f})   &  p_{N+N_f}(z)\\
K_{N+1}(m_1,z^*) & \ldots\ K_{N+1}(m_{N_f},z^*) & K_{N+1}(z,z^*)\\
\end{array} 
\right| ,
\label{ZNNfn+1}
\ee
where $K_{N+1}(x,y)$ is the bare kernel introduced in (\ref{Kdef}).
In the replica limit of the Toda lattice equation, the
partition function with one pair of conjugate bosons instead of
conjugate fermions enters as well. 
It is given by \cite{BII} (see also \ref{app:Z})
\be
\label{ZNNfn-1}
\frac{{\cal Z}_{N}^{N_f,n=-1}(\{m_f\}|y,x^*;\mu)}{{\cal Z}_{N}^{N_f=0}}  
&\equiv&
\left\langle\ \frac{\prod_{f=1}^{N_f}m_f^\nu}{(yx^*)^\nu}\prod_{j=1}^N \left(\frac{1}{(z_j^2-y^2)(z_j^{*\, 2}-x^{*\,2})}
\prod_{f=1}^{N_f}
(m_f^2-z_j^2 )\right)\ \right\rangle_{N_f=0} \\
&&\hspace*{-5cm}
=
-\frac{\prod_{f=1}^{N_f}m_f^\nu(m_f^2-y^2)}{r_{N-1}\ 
\Delta_{N_f}(\{m_f^2\})(yx^*)^\nu}
\left|
\begin{array}{lll}
p_{N-1}(m_1)     & \ldots\ p_{N-1}(m_{N_f})  & h_{N-1}(x^*)  \\
\hspace{5mm}\vdots & \ddots\ \hspace{5mm}\vdots & \hspace{5mm}\vdots\\
p_{N+N_f-2}(m_{1}) & \ldots\ p_{N+N_f-2}(m_{N_f})& h_{N+N_f-2}(x^*)\\
{\cal N}_{N+N_f-2}(m_1,y)   & \ldots\ {\cal N}_{N+N_f-2}(m_{N_f},y)&
{\cal A}_{N+N_f-2}(x^*,y)
\end{array}
\right|~.\nn
\ee
If we take the limit of degenerate bosonic masses, $y\to x$, the
function $Q(x^*,y;\mu)$ inside the matrix element 
${\cal A}_{N+N_f-2}(x^*,y)$ becomes singular, while all other matrix
elements remain finite. Expanding the determinant with respect to the last row
this singular part $\sim Q_\epsilon(x^*,x;\mu)$ 
gives the dominant contribution, 
\be
\frac{{\cal Z}_N^{N_f,n=-1}(\{m_f\}|x,x^*;\mu)}
{{\cal Z}_N^{N_f=0}}  
&\stackrel{\epsilon\to0}{\sim}& -
\frac{\prod_{f=1}^{N_f}m_f^\nu(m_f^2-x^2)}{r_{N-1}\ 
\Delta_{N_f}(\{m_f^2\})|x|^{2\nu}}                           
\det_{1\leq k,l\leq N_f}\left[\ p_{N+k-2}(m_l)\ \right]\ 
{\cal A}_{N+N_f-2}(x^*,x) ~.
\label{ZNF1n-1fac}
\ee
Using the relations (\ref{Adef}) and (\ref{ZNf}) we thus arrive at 
the following factorized result,
\be
\lim_{\epsilon\to 0}
\frac{{\cal Z}_N^{N_f,n=-1}(\{m_f\}|x,x^*;\mu)}{{\cal Z}_N^{N_f=0}}  
&=& 
\prod_{f=1}^{N_f}(m_f^2-x^2)             
\frac{{\cal Z}_{N-1}^{N_f}(\{m_f\};\mu)}{{\cal Z}_{N-1}^{N_f=0}}\ 
\lim_{\epsilon \to 0}\frac{{\cal Z}_N^{n=-1}(x,x^*;\mu)}{{\cal Z}_{N}^{N_f=0}}\ .
\label{ZNF1n-1log}
\ee
Both sides are regularized by replacing $Q(x,x^*;\mu)$ by 
$Q_\epsilon(x,x^*;\mu)$  as discussed above (\ref{Qeps}).
This is the main result of this section. 
Let us stress that the factorization we obtained is not a consequence of the
suppression of non-planar diagrams in the large $N$ limit. 
In (\ref{ZNF1n-1log}) it occurs at finite-$N$ and 
is strictly due to the singularity for coinciding arguments.
This completes the derivation of all generating functions 
required for the derivation of the spectral
density from the replica limit of the Toda lattice equation (\ref{rhoToda}).

If we were to compute not only the spectral density but also higher order
correlation functions from the Toda lattice approach (see e.g. in 
\cite{SplitVerb2}), 
we would have to add more pairs of replicated bosons, each with
different masses $y_j$ and $x_j^*$. Taking the degenerate limit $y_j\to x_j$
would lead to the same type of factorization into
$\sim \prod_j  {\cal  Z}_N^{n_j=-1}(x_j,x_j^*;\mu)$  
and a remaining fermionic partition function.

%%%%%%%%%%%%%%%%%%%%%%%%%%%%%%%%%%%%%%%%%%%%%%%%%%%%%%%%%%%%%%%%%%%%%%%%%%
\section{Toda lattice equation from orthogonal polynomials}
\label{todaop}

The purpose of this section is threefold. First we show that the 
generating functions with $n$ pairs of fermionic replicas satisfy the 
Toda lattice equation. Since the microscopic  limit 
of the random matrix partition function,
${\cal Z}_{N}^{N_f,n}$, is given by the effective 
partition function (\ref{zeff}) \cite{SplitVerb2,AFV}, this result 
also shows that the Toda lattice equation (\ref{TodaBaryon})
for $N_f = 0$ \cite{SplitVerb2} can be extended to arbitrary $N_f$.
Second, we show that the spectral density obtained from
the replica limit of the Toda Lattice equation agrees with the result
computed from orthogonal polynomials \cite{O}. This extends 
the agreement between the two approaches to cases where the weight in
the partition function is no longer positive definite.
We close this section with some remarks about universality.

\subsection{The spectral density from the Toda lattice equation}

We start by showing that the Toda lattice equation 
(\ref{TodaBaryon}) can be modified to hold for
random matrix model partition functions at finite $N$. 
The starting point of the evaluation is the expression for the
partition function with $N_f+n$ ordinary flavors and $n$ conjugate fermionic 
flavors generalizing (\ref{ZNNfn+1}). 
While this result is given in \cite{AV} for non-degenerate masses we
have to take the limit where the replicated masses become degenerate. 
This leads to successive derivatives in these variables of the polynomials and
of the kernels (\ref{Kdef}). 
For simplicity we only display the result for $N_f=1$, 
\be
\frac{{\cal Z}_{N}^{N_f=1,n}(m,z,z^*;\mu)}{{\cal Z}_{N}^{N_f=0}} 
& \sim & \frac{m^\nu |z|^{2n\nu}\prod_{j=N}^{N+n-1}r_j }{(m^2-z^2)^n 
|z|^{n(n-1)}}
\\
&&\hspace{-45mm}
\times \left|\begin{array}{ccccc} 
\np(m) & \np(z) & \delta_z \np(z) 
& \cdots & \delta_z^{n-1} \np(z)\\
\nK(m,z^*) & \nK(z,z^*) & \delta_z \nK(z,z^*) 
& \cdots & \delta_z^{n-1}\nK(z,z^*) \\
\delta_{z^*}\nK(m,z^*) & \delta_{z^*}\nK(z,z^*) & \delta_{z^*}
\delta_z \nK(z,z^*) &\cdots & \delta_{z^*}\delta_z^{n-1}\nK(z,z^*) \\
\vdots & \vdots & \vdots 
& \ddots & \vdots \\
\delta_{z^*}^{n-1}\nK(m,z^*) & \delta_{z^*}^{n-1}\nK(z,z^*) 
& \delta_{z^*}^{n-1}\delta_z \nK(z,z^*) 
& \cdots & \delta_{z^*}^{n-1}\delta_z^{n-1}\nK(z,z^*) 
\end{array}\right|\nn
\ee
omitting constant factors. Here the derivatives are defined as
\be
\delta_z = z \frac d{dz}, \qquad \delta_{z^*} = z^*\frac d{dz^*}.
\ee
We observe that this expression 
depends on the matrix size explicitly through the combination
$N+n$ and implicitly through factors that originate from the $N$-dependence
in the exponent of the weight function. 
Using the Sylvester identity as
in \cite{SplitVerb2} while  
keeping only the $N$ dependence that enters as $N+n$
we easily derive the Toda lattice equation,
\be 
\delta_z\delta_{z^*} \log {\cal Z}_{N}^{N_f=1,n}(m,z,z^*;\mu) 
& \sim & n(zz^*)^2
\frac{{\cal Z}_{N-1}^{N_f=1,n+1}(m,z,z^*;\mu)
{\cal Z}_{N+1}^{N_f=1,n-1}(m,z,z^*;\mu)}
{[{\cal Z}_{N}^{N_f=1,n}(m,z,z^*;\mu)]^2}\ ,
\ee
again ignoring constant factors. This equation is valid at finite $N$
provided that the factor $N$ that appears explicitly in the exponent
of the weight function is
not changed. The subscript of the partition function thus denotes the
total number of complex integration variables. 
For large  $N$ this distinction can be ignored.
It is not difficult to generalize the Toda lattice equation to arbitrary
$N_f$. The result is  
\be 
\label{TodaBaryonN}
\delta_z\delta_{z^*} \log {\cal Z}_{N}^{N_f,n}(\{m_f\},z,z^*;\mu) 
&\sim&  n(zz^*)^2
\frac{{\cal Z}_{N-1}^{N_f,n+1}(\{m_f\},z,z^*;\mu)
{\cal Z}_{N+1}^{N_f,n-1}(\{m_f\},z,z^*;\mu)}
{[{\cal Z}_{N}^{N_f,n}(\{m_f\},z,z^*;\mu)]^2}\ . 
\ee 
We still have to fix the overall normalization constant
of the partition functions which in principle depends on $N,\ N_f$, as well 
as on 
the replica index $n$. Since the l.h.s. of (\ref{TodaBaryonN}) is linear
in $n$ we have to  choose the normalization constants in the r.h.s. 
to retain this $n$-dependence. With the appropriately adjusted 
constants the random matrix model generating functions
satisfy the
equality   (\ref{TodaBaryon}).

Since the microscopic limit of the random  matrix partition functions 
${\cal Z}_{N}^{N_f,n}(\{m_f\},z,z^*;\mu) $ 
is given by the group integrals (\ref{zeff}), 
we conclude that (\ref{TodaBaryon}) is satisfied  for
any number of flavors. 
Explicit expressions for these  
partition functions in terms of Bessel functions will be given in
section \ref{results} where we also give results for 
the spectral densities.

Before doing so let us step back to discuss what we have achieved for the
spectral density using the replica limit of the Toda lattice equation
(\ref{TodaBaryonN}).
Combining equations (\ref{TodaBaryon}) and (\ref{replica})
with the factorization (\ref{ZNF1n-1log}) and canceling one of
the partition functions in the denominator of the Toda lattice equation,
we arrive at the 
following result for the spectral density, 
\be
\rho^{N_f}_\nu(z,z^*,\{m_f\};\mu) 
& = &  - \frac{2}{\pi \log{\epsilon}}\frac {zz^*}2 
\prod_{f=1}^{N_f}(m_f^2-z^2) 
{\cal Z}^{n=-1}_{N+1}(z,z^*;\mu)
\frac{{\cal Z}^{N_f,n=1}_{N-1}(\{m_f\},z,z^*;\mu)}
{{\cal Z}^{N_f}_N(\{m_f\};\mu)}\ ,
\label{rhoNfN}
\ee
where we have explicitly displayed
the divergent normalization factor $-(\pi/ 2) \log \epsilon$  which
cancels the divergent factor that was obtained in (\ref{QZ}). 
The $\nu$ dependence of
the spectral density is contained implicitly in the partition functions.
This is the central result of the Toda lattice approach 
applied to the random matrix model (\ref{ZNfNb}). 
The spectral density factorizes into QCD like partition functions
which, at low energy, can be expressed as group integrals of the form
(\ref{zeff}) and (\ref{ZMINQCD}). 

Let us discuss some properties of the spectral density. 
First, it is symmetric under 
$z\to-z$ (together with $z^*\to-z^*$), because the partition function
${\cal Z}^{N_f,n=1}_N$ contains only squared variables and
${\cal Z}^{n=-1}_N$ is even as well. This is a consequence of the
axial symmetry of the partition function which requires that the nonzero
eigenvalues occur in pairs of opposite sign.  
Inserting the result (\ref{ZNNfn+1}) for ${\cal Z}^{N_f,n=1}_N$ 
we see that the prefactor $\prod_{f=1}^{N_f}(m^2_f-z^2)$ is canceled from 
the Vandermonde (for an explicit expression in terms of determinants see 
(\ref{rhokernel}) below). Replacing the polynomials
and kernels in ${\cal Z}^{N_f,n=1}_N$ 
by their expressions in terms of partition functions
we notice one further symmetry 
(for real quark masses), 
\be
\rho^{N_f}_\nu(z,z^*,\{m_f\};\mu)^{\,*}\ =\ \rho^{N_f}_\nu(z^*,z,\{m_f\};\mu)\ ,
\label{rhosym}
\ee
that is taking the complex conjugate and at the same time exchanging $z$ and
$z^*$. These symmetries follow directly if we write the spectral density as
an integral over the joint eigenvalue distribution. 
Thus the real part is symmetric about the real axis in the complex
eigenvalue plane while the imaginary part is anti-symmetric. 
Therefore, the eigenvalue density on the real axis 
is real. However, away from the real axis 
it is, in general, neither real nor positive and therefore 
does not define a probability density.
The lack of reality directly relates to the asymmetric way the variables 
$z$ and $z^*$ enter in the polynomials and kernel respectively.
%It is most likely that these symmetries are
The conjugation symmetry is related to the invariance of the partition function
under changing the sign of $\mu$, and is
valid for QCD beyond the ergodic domain (\ref{epslim}) as well.
%jv

\subsection{Equivalence of the Toda lattice and 
orthogonal polynomials approach}
%\label{app:rhofiniteN}

In order to compare the result (\ref{rhoNfN}) from the replica
limit of  the Toda lattice equation
to the result from orthogonal polynomials \cite{O} it is 
instructive to rewrite (\ref{rhoNfN})  in terms of the weight function,
polynomials and kernels. Reinserting (\ref{QZ}), (\ref{ZNNfn+1}) 
and (\ref{ZNf}) we obtain
\be
\rho^{N_f}_\nu(z,z^*;\{m_f\}, \mu) \ =\ 
w(z,z^*;\mu)\ 
\frac{\left|
\begin{array}{lll}
p_{N}(m_1)     & \ldots\ p_{N}(m_{N_f})     & p_{N}(z)\\
~~\vdots & \ddots~~~\vdots & ~~\vdots\\
p_{N+N_f-1}(m_1)   & \ldots\ p_{N+N_f-1}(m_{N_f})   &  p_{N+N_f-1}(z)\\
K_{N}(m_1,z^*) & \ldots\ K_{N}(m_{N_f},z^*) & K_{N}(z,z^*)\\
\end{array}
\right|}{\det_{1\leq k,l\leq N_f}\left[\ p_{N+k-1}(m_l)\ \right]}\ ,
\label{rhokernel}
\ee
again after normalizing appropriately. 
As we will explain next,
this is precisely the result that
was obtained 
in \cite{O}. 

It is well-known from random matrix theory that the spectral density can be 
written in terms of a kernel of orthogonal polynomials 
\cite{Mehta}. This also holds for complex weight
functions. However, if the weight function is not a symmetric function
of $z$ and $z^*$, we have to use
bi-orthogonal polynomials 
in the complex plane defined by \cite{BI,BII},
\be 
\int_{\mathbf{C}} d^2z\ w^{N_f,0}_\nu(z,z^*;\{ m_f\}, \mu)  \
p^{(N_f)}_k(z) \ q_l^{(N_f)}(z^*) \ =\ r_k^{(N_f)}\delta_{kl} \ ,
\label{biOPdef}
\ee 
assuming the pseudo norms $r_k^{(N_f)}$ to be nonvanishing. 
(Bi-orthogonal polynomials in general no longer form a scalar product with 
positive definite norms.)
In contrast to orthogonal polynomials on the real line, the 
$p_k(z)$ and $q_k(z)$ are in general different polynomials.
Defining their  kernel by
\be
{\cal K}_N^{(N_f)}(z,u^*)
\ \equiv\ [w^{N_f,0}_\nu(z,z^*;\{m_f\},\mu)w^{N_f,0}_\nu(u,u^*;\{m_f\},\mu) 
%\prod_{f=1}^{N_f}(z^2+m_f^2)(u^{*\,2}+m_f^2)
]^{\frac12}
\sum_{j=0}^{N-1} \frac{1}{r_j^{(N_f)}}\ p^{(N_f)}_j(z) q_j^{(N_f)}(u^*) ~,
\label{biKdef}
\ee
the following expression for the correlator of $k$ complex 
eigenvalues holds \cite{Mehta}
\be
\rho^{N_f}_\nu(\{z_l,z^*_l\}_{l=1,\ldots,k};\{m_f\};\mu)\ =\ 
\det_{1\leq j,l\leq k}\left[\ {\cal K}_N^{(N_f)}(z_j,z_l^*)\ \right].
\label{rhok}
\ee
In particular, for the spectral density we obtain
\be
\rho^{N_f}_\nu(z,z^*;\{m_f\};\mu)\ =\ w^{N_f,0}_\nu(z,z^*;\{m_f\};\mu) 
%\prod_{f=1}^{N_f}(z^2+m_f^2)
\sum_{j=0}^{N-1} \frac{1}{r_j^{(N_f)}}\ p^{(N_f)}_j(z) q_j^{(N_f)}(z^*). 
\label{rhobiOP}
\ee
In \cite{O} it was shown that the kernel for $N_f$ flavors can be
expressed as a determinant of kernels for zero flavors
which immediately leads to the result (\ref{rhokernel}).

This establishes the equivalence between the Toda lattice approach
and the bi-orthogonal
polynomial approach at the level of the spectral density. 
Starting from the observation that also in the case of 
generating functions 
for  multi-point correlation functions the partition function 
 ${\cal Z}_N^{n=-1}(z_l,z_l^*;\mu)$  factors out,
we are confident that the
equivalence between the Toda lattice approach and the bi-orthogonal 
polynomial approach can be established in that case as well.

\subsection{Universality}
%\label{sec:microdens}

After having computed the spectral density from the replica limit
of the Toda lattice equation and compared to that from the
orthogonal polynomial approach,
we would like to use the insight gained to discuss the issue of universality.
First of all, we can distinguish between two different ways to 
approach universality. The first way is based on the observation 
that different theories with the same pattern of spontaneous symmetry
breaking and a mass gap have the same low-energy limit. For example,
gauge field theories and random matrix theories with the same
global symmetries are described by the same effective partition function in the
microscopic scaling limit. For QCD with three colors and fundamental
quarks at nonzero chemical potential this universal partition function
is given by the group integral (\ref{zeff}). In QCD, a mass gap exists
because of confinement, and in Random Matrix Theory,  a mass gap appears
in the large $N$ limit. Therefore, if we change the action 
without affecting the global symmetries and the 
existence of a mass gap, the microscopic scaling limit will remain
the same. A second way to understand  universality  is based on a
direct calculation of correlation functions for a deformed probability
distribution which does not necessarily have the unitary invariance
of the Gaussian model.
In the case of deformations that respect the unitary
invariance,  universality then follows from
the asymptotic behavior of the modified orthogonal polynomials.
In particular, for chiral random matrix theory at $\mu =0$ this
approach has been very successful \cite{FK,ADMN}.

Let us also comment on the role of the weight function in 
non-Hermitian random matrix theories. 
From the replica limit of the Toda lattice equation it is clear
that the eigenvalue density will be proportional to the partition
function with one pair of conjugate bosonic flavors which,
according to this work, is proportional to the weight function. 
In general, this weight function contains both universal and non-universal 
parts, and in the microscopic scaling limit only the universal parts
survive.
If we allow for deformations of the Gaussian weight by higher order 
powers the resulting weight function, 
and therefore the spectral density, will be modified. 
It is only in the microscopic scaling  
limit that the non-universal parts of the weight function disappear and
a universal spectral density will be recovered. 
In Hermitian matrix models the weight function at finite $N$ is also 
non-universal
but reduces to a universal result in the microscopic scaling limit. 
Non-Hermitian random matrix models for QCD at $\mu\neq0$ represent a
continuous one parameter deformation of the Hermitian model at $\mu=0$,
governed by $\mu^2N$. In this case, the 
weight function contains an additional universal piece in the 
microscopic scaling limit  that develops into a delta function 
$\sim\delta({\rm Im}(z))$ for  $\mu^2N\to0$. The additional universal
piece occurs because an eigenvalue representation can only be obtained
after a nontrivial integration over the similarity transformations that
diagonalize the  non-Hermitian matrices. In the Hermitian case it is trivial to
obtain the eigenvalue representation, and no additional universal piece
is generated. For technical reasons we have not been able to integrate
out the similarity transformations when higher powers are added to the
Gaussian probability distribution of the model (\ref{ZNfNb}).
We mention in passing that for non-chiral non-Hermitian random matrix models a 
universality proof by deforming the random matrix potential could be 
given \cite{A02}.

%%%%%%%%%%%%%%%%%%%%%%%%%%%%%%%%%%%%%%%%%%%%%%%%%%%%%%%%%%%%%%%%%%%%%%%%%%%%%%
\section{Results and explicit examples}
%\label{sec:Examples}
\label{results}

\subsection{The partition functions in the microscopic limit}
\label{sec:micro}

Having obtained analytical expressions for 
the partition functions of the
random matrix model (\ref{ZNfNb}) at finite $N$ 
we now consider the large $N$ limit. This
limit will be taken according to the counting scheme where $m_fN$ and $\mu^2N$
are kept fixed as $N\to\infty$ and is referred to as the microscopic limit.
The rescaling of the non-Hermiticity parameter $\mu^2$ with $N$ was first
introduced in \cite{FS} as the concept of weak non-Hermiticity.
As discussed in the previous section,
in the microscopic limit the random matrix partition function is given
by an  integral over the Goldstone manifold, i.e. it
is uniquely determined by the spontaneous breaking of the flavor symmetries. 
The random matrix model can be viewed as an alternative way to calculate the
integral over the Goldstone manifold, which in the case of nonzero 
chemical potential and both bosonic and fermionic quarks has not
been accomplished in any other way.

The dependence of the quark mass and the chemical potential in the integral 
over the
Goldstone manifold is only through the combinations $m \Sigma V$ and $\mu^2
F_\pi^2 V$, cf. (\ref{zeff}) and (\ref{ZMINQCD}) \cite{dominique-JV}.  
The dimensionful scales can be recovered from the random matrix model by 
making the replacements
\be
2m N & \to & m V \Sigma \\
2\mu^2 N & \to & \mu^2 F_\pi^2 V. \nn
\ee
To make contact with the physical content of the theory we will express
our results in terms of the dimensionful scales 
in the remainder of this section. 

\vspace{3mm}

In the case we have only either fermionic or bosonic quarks and no
conjugate quarks, the Goldstone bosons are not charged with respect to
the chemical potential, and we should find the microscopic partition
functions at zero chemical potential. One easily observes that
the effective partition
function (\ref{zeff}) does not depend on the chemical potential
in this case.
In the microscopic limit of (\ref{Heine}) we thus find the partition
function (\ref{zeff}) with one fermionic flavor \cite{GL,KSS,Brower,jsv}
\be
Z^{N_f=1}_\nu(m;\mu) = I_\nu(m \Sigma V).
\label{I}
\ee
An overall scaling factor $\mbox{e}^{-\mu^2 F_{\pi}^2 V}$ has been removed. 
This unphysical factor can be removed in 
the random matrix model by scaling the
Gaussian potential in the weight (\ref{wg}) by a factor of $1-\mu^2$ \cite{O}.
We did not include an explicit scale factor in the weight function, but will
assume this procedure has been done in order 
to arrive at the correct expressions in
the microscopic limit.

%The microscopic partition function with one bosonic flavor is obtained from 
%(\ref{CHeine}). It is given by the Cauchy transform of the microscopic
%partition function of one fermionic flavor which is the  modified
%Bessel function $K_\nu$, 
%\be
%Z^{N_f=-1}_\nu(m;\mu) = K_\nu(m \Sigma V).
%\ee
%Since there are no charged Goldstone modes this partition  
%correctly  coincides with the $\mu=0$ result \cite{DV}. 

When there is one pair of conjugate fermionic quarks, the microscopic
partition function is given by the
result obtained by performing the integral (\ref{zeff}) over the Goldstone
manifold \cite{SplitVerb2}
\be
Z^{n=1}_\nu(y,z^*;\mu)  = \mbox{e}^{2\mu^2F_\pi^2 V}
\int_0^1 dt \ t \ \mbox{e}^{-2\mu^2F_\pi^2 V t^2}
I_\nu(y \Sigma V t)I_\nu(z^*\Sigma V t).
\label{zn=1}
\ee
For the universal partition function with a pair of conjugate bosonic quarks we
find using (\ref{QZ})
\be
Z^{n=-1}_\nu(z,z^*;\mu)  = (|z|\Sigma V)^{2\nu} 
\frac{1}{\mu^2F_\pi^2 V}
\exp\left(- \frac{(z^2 + \conj{z}^2)\Sigma^2V}{8 \mu^2F_\pi^2}  \right)
K_\nu \left( \frac{|z|^2\Sigma^2V}{4 \mu^2F_\pi^2}  \right),
\ee
where we have divided out the divergent factor $-\pi\log(\epsilon)/2$.
This is again in agreement with the result \cite{SplitVerb2} obtained by
an explicit integration over the Goldstone manifold (\ref{ZMINQCD}).

All other microscopic partition functions 
required to calculate the spectral density 
can be obtained from
the relation (\ref{ZNF1n-1log})
\be
Z^{N_f,n=-1}_\nu(\{m_f\},z,z^*;\mu)=\prod_{f=1}^{N_f}(m_f^2-z^2)
     Z^{N_f}_\nu(\{m_f\};\mu)Z^{n=-1}_\nu(z,z^*;\mu).
\label{ZNfn=-1}
\ee

\begin{figure}[!ht]
  \unitlength1.0cm
  \begin{center}
  \begin{picture}(3.0,2.0)
  \put(-5.,-5.){
  \psfig{file=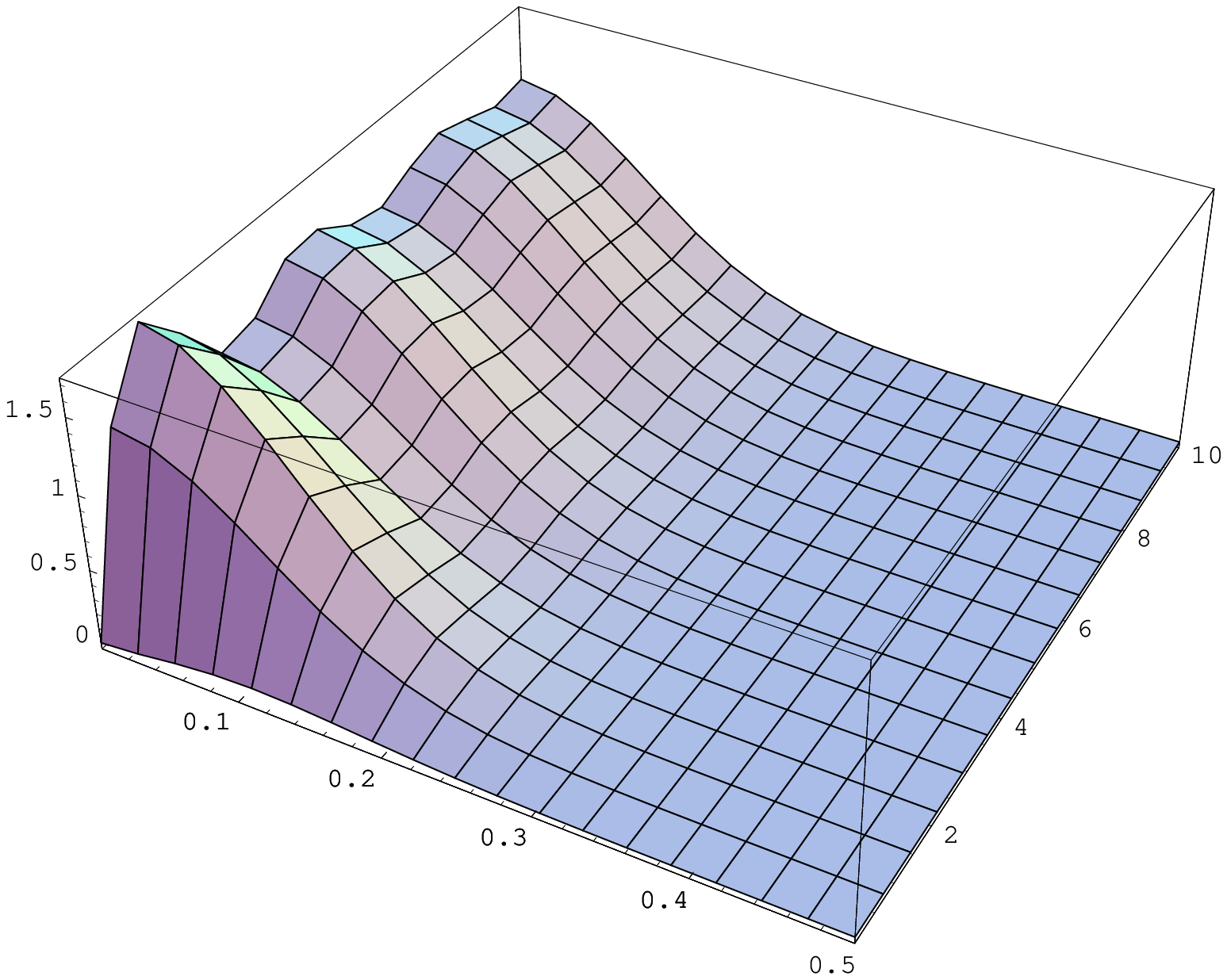,clip=,width=10cm}}
  \put(-1.5,-4.9){\bf\large Re$[z]\Sigma V$}
  \put(4.5,-2.0){\bf\large Im$[z]\Sigma V$}
  \put(-6.8,1.7){\bf \LARGE
$\frac{\rho^{N_f=0}_{\nu=0}(z,z^*;\mu)}{\Sigma^2V^2}$}
  \put(-5.,-14.){
  \psfig{file=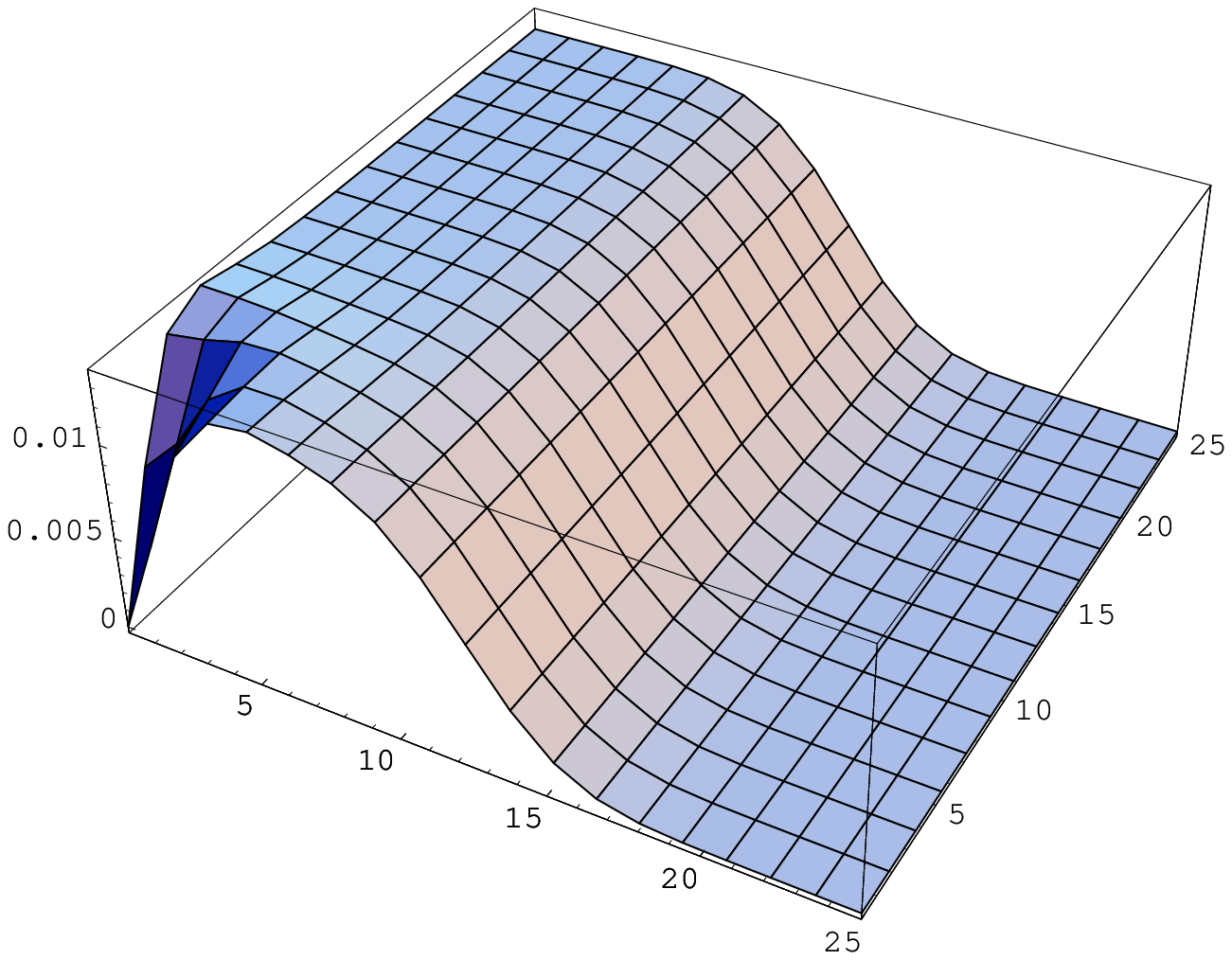,clip=,width=10cm}}
  \put(-1.3,-13.8){\bf\large Re$[z]\Sigma V$}
  \put(5,-10.4){\bf\large Im$[z]\Sigma V$}
  \put(-6.8,-7.4){\bf \LARGE$\frac{\rho^{N_f=0}_{\nu=0}(z,z^*;\mu)}{\Sigma^2V^2}$}
  \end{picture}
  \vspace{14cm}
  \end{center}
\caption{\label{fig:rhoQmu01} The quenched
  eigenvalue density with a chemical potential
$\mu F_\pi \sqrt{V}=0.1$ (upper) and $\mu F_\pi \sqrt{V}=2.5 $ (lower). 
  The density is real and positive and  
 follows in the other
  quadrants by reflection on the axes. In the upper figure
 $\mu F_\pi \sqrt{V}\ll1$, and one can
  still see the individual eigenvalue distributions. In the lower figure
  the
  repulsion of eigenvalues from the origin is apparent. The width of the strip
  is on the order Re$[z]\Sigma\sim 2 \mu^2F^2_\pi $. 
The normalization is such that the
  width times the hight is independent of $\mu$.}   
\end{figure}

\subsection{The microscopic spectral density}

In this section we write out several different examples for the eigenvalue
density of the QCD Dirac operator in the microscopic limit. We emphasize
that these results are nonperturbative analytical predictions for 
the QCD Dirac spectrum at nonzero chemical potential
which should be reproduced by lattice QCD simulations. In order to
get more compact expressions we will express our results 
in terms of the partition function (\ref{zn=1}).
The normalization of the quenched density is chosen 
such that there are $\Sigma V/\pi$ eigenvalues per unit length   
along the imaginary axis. The density of the projection of the eigenvalues
on the imaginary axis is therefore equal to the eigenvalue density at
$\mu = 0$.
The other densities reduce to the quenched density
when the quark masses are taken to infinity ($m_f\Sigma V \gg \mu^2F_\pi^2 V$) 
and we use this limit to fix the normalization. 

The general expression for the spectral density is given by
\be
\rho^{N_f}_\nu(z,z^*,\{m_f\};\mu) 
& = &  \frac{zz^*}2 
\prod_{f=1}^{N_f}(m_f^2-z^2)
Z^{n=-1}_\nu(z,z^*;\mu)
\frac{Z^{N_f,n=1}_\nu(\{m_f\},z,z^*;\mu)}
{Z^{N_f}_\nu(\{m_f\};\mu)}\ ,
\label{rhofinal}
\ee
where the microscopic limit of ${\cal Z}^{n=-1}_N(z,z^*;\mu)$ is given in the
previous subsection. The microscopic limit of ${\cal Z}^{N_f}_N(\{m_f\};\mu)$
is given by
\be
Z^{N_f}_\nu(\{m_f\};\mu) = \frac{{\rm det}
[(x_k\del_{x_k})^l I_\nu(x_k)]_{k,l=0,\cdots,N_f-1}}{\Delta(x_k^2)},
\ee
where $x_k = m_k V\Sigma$.
Note that with only regular fermionic quarks, $\mu$ has no effect on the
microscopic partition function.
The partition function 
${\cal Z}^{N_f,n=1}_N(\{m_f\},z,z^*;\mu)$ is given in terms of the polynomials
$p_N(m)$ and the bare kernel $K_{N+1}(x,y)$ in (\ref{ZNNfn+1}). 
The microscopic limit of this partition function is obtained by expressing
the polynomials and the kernel in terms of partition functions that
are known in the microscopic limit. The polynomials
$p_N(m)$ can be expressed 
as the partition functions with one fermionic flavor (see (\ref{Heine})),
which in the microscopic limit does not depend on $\mu$ and is given 
by $I_\nu(mV\Sigma)$ in (\ref{I}). In (\ref{KZ}) the kernel  $K_{N+1}(x,y)$
was shown to be equal to $\tK$ which  is given in (\ref{zn=1}).
In the remaining sections we will set $\nu=0$ for simplicity.

\subsubsection{The quenched spectral density}

The quenched microscopic eigenvalue density of the Dirac operator at nonzero
baryon chemical potential is given by \cite{SplitVerb2} 
\be\label{rhoquenched}
\rho^{N_f=0}_{\nu = 0}(z,z^*;\mu)
 &=&  \frac{|z|^2\Sigma^4V^3}{2\pi\mu^2F_\pi^2}\mbox{e}^{-2\mu^2 F_\pi^2 V}
\mbox{e}^{-\frac{(z^2+z^{*\,2})\Sigma^2 V}{8\mu^2F_\pi^2}} 
K_0\left(\frac{|z|^2\Sigma^2 V}
{4\mu^2F_\pi^2}\right)Z^{n=1}_{\nu=0}(z,z^*;\mu).
\ee
This result has been checked by direct numerical simulations
\cite{SplitVerb2} within a random matrix model, and successfully 
compared to results from lattice QCD \cite{GT}. Plots of the quenched 
spectral density are given in Fig. \ref{fig:rhoQmu01}. 

By using the leading order asymptotic forms of the Bessel functions, one
easily shows that
in the limit $\mu^2F_\pi^2V \gg 1$ the support of the quenched spectrum is
inside the strip \cite{SplitVerb2,minn04}
\be
\label{zstrip}
\frac {|\mathrm{Re}(z)|\Sigma}{2\mu^2F_\pi^2} < 1
\ee
with a density that is equal to
\be
\rho^{N_f=0}_{\nu=0}(z,z^*;\mu) = \frac {V\Sigma^2}{4\pi\mu^2F_\pi^2}.
\ee
In Fig. \ref{fig:rhoQmu01}
we observe that this plateau is already present for 
$\mu F_\pi \sqrt V=2.5$.  This step in the spectral density 
is associated with a phase transition of the generating function. 
 The theory with a pair of conjugate fermionic flavors of mass
$x$, described by $Z_{n=1}(x,x;\mu)$, is in the normal phase for
$x\Sigma/F_\pi^2=m_\pi^2/2>2\mu^2$ and in a Bose condensed phase for 
$x\Sigma/F_\pi^2=m_\pi^2/2<2\mu^2$ \cite{dominique-JV}.
For the $x$-integrated quenched spectral density we find ($z= x+iy$)
\be
\int_{-\infty}^\infty dx\,\rho_{\nu=0}^{N_f=0}(z,z^*;\mu) = \frac{\Sigma V}\pi,
\ee
consistent with our choice of normalization.

\subsubsection{The spectral density for one flavor}

\begin{figure}[!ht]
  \unitlength1.0cm
  \begin{center}
  \begin{picture}(3.0,2.0)
  \put(-5,-5){
  \psfig{file=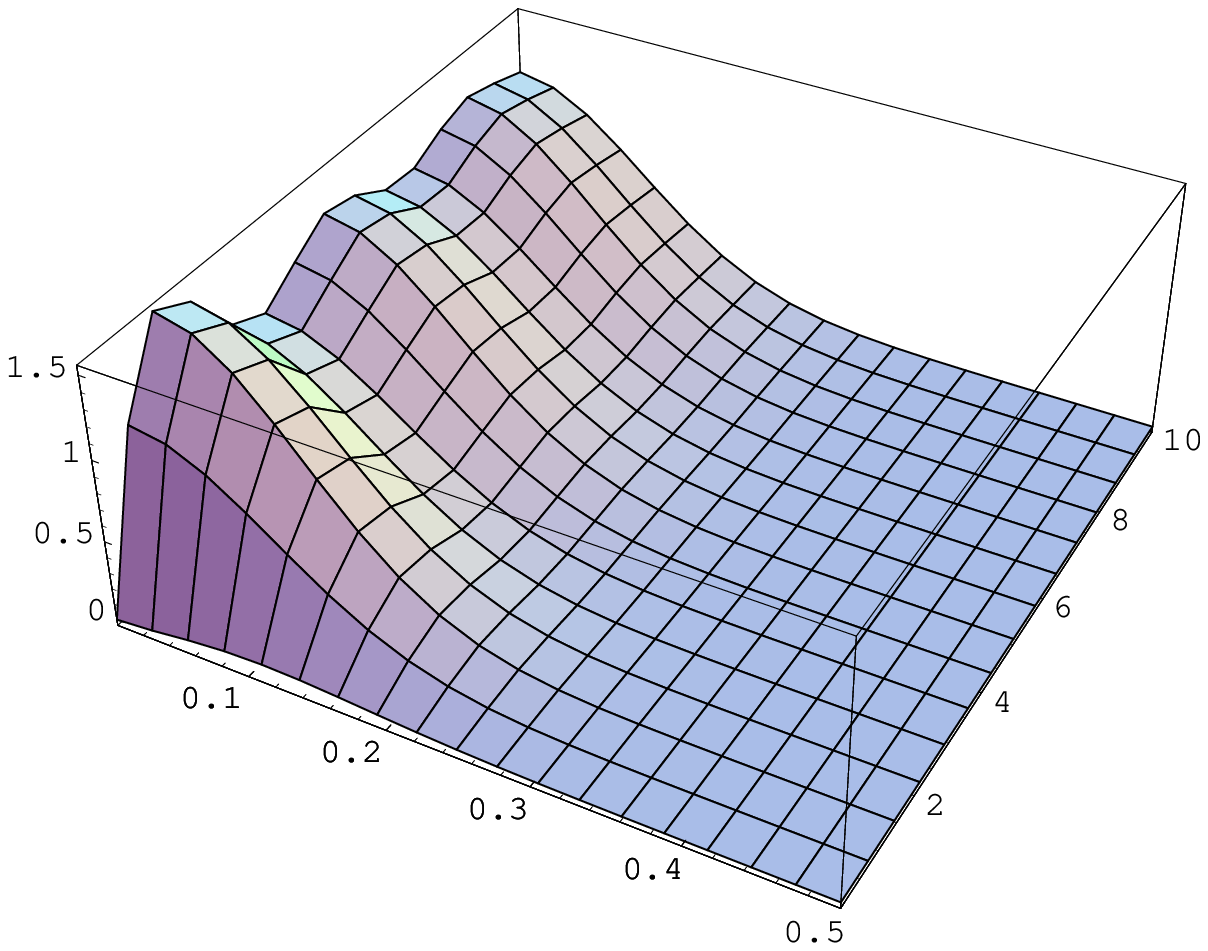,clip=,width=10cm}}
  \put(-1.3,-5.0){\bf\large Re$[z]\Sigma V$}
  \put(5,-1.4){\bf\large Im$[z]\Sigma V$}
  \put(-6.8,2.1){\bf \LARGE $\frac{{\rm Re}[\rho^{N_f=1}_{\nu=0}(z,z^*,m;\mu)]}{\Sigma^2V^2}$}
  \put(-5.,-14){
  \psfig{file=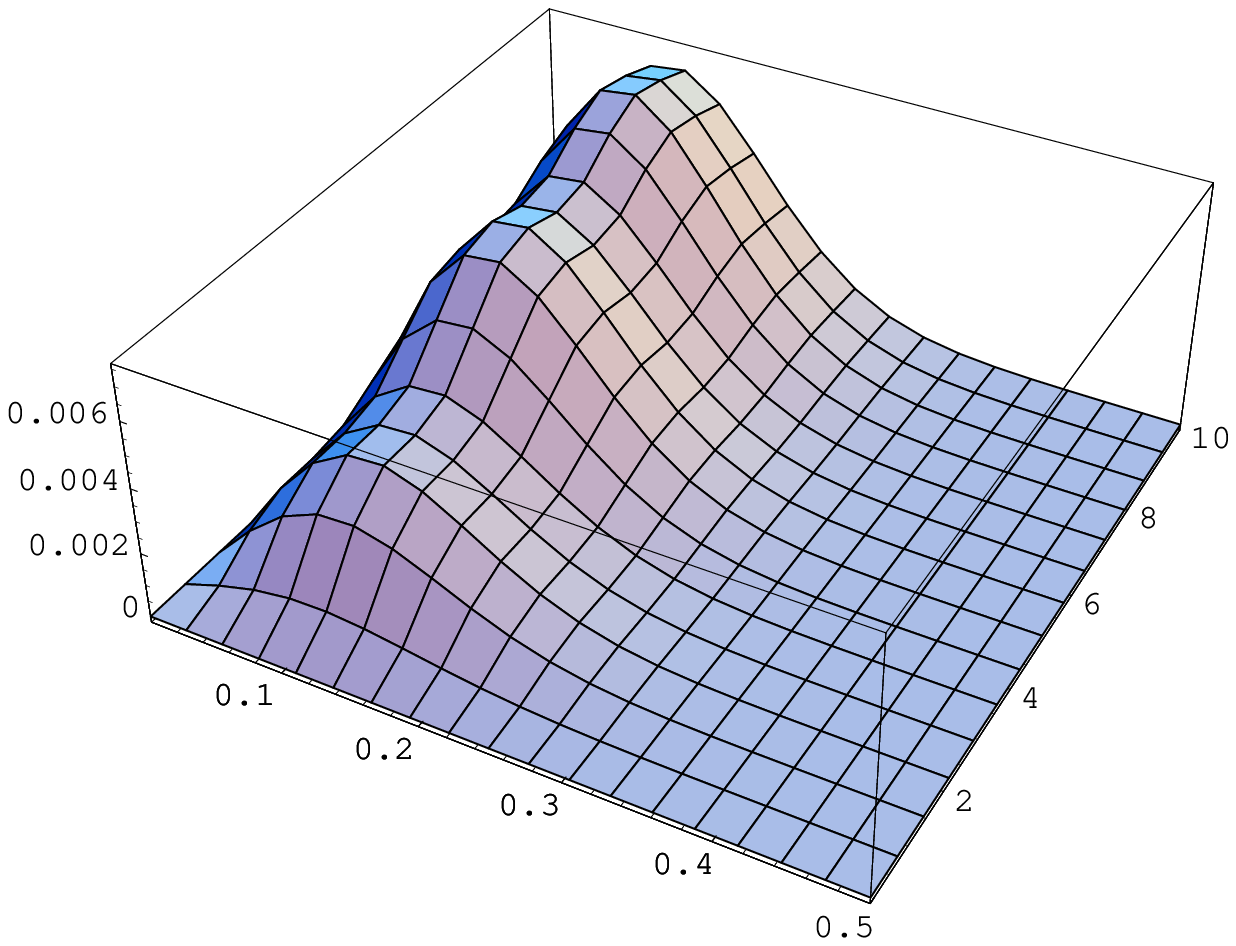,clip=,width=10cm}}
  \put(-1.3,-13.8){\bf\large Re$[z]\Sigma V$}
  \put(5,-10.4){\bf\large Im$[z]\Sigma V$}
  \put(-6.8,-7.2){\bf \LARGE $-\frac{{\rm Im}[\rho^{N_f=1}_{\nu = 0}(z,z^*,m;\mu)]}{\Sigma^2V^2}$}
  \end{picture}
  \vspace{14cm}
  \end{center}
\caption{\label{fig:Rerhoa1m10mu01} Real and imaginary parts 
of the eigenvalue density for one 
  flavor of mass $mV\Sigma=10$ and chemical potential 
$\mu F_\pi\sqrt V=0.1$. 
With these values the real part of the density is quite similar
  to the quenched density.
Note that the scale of the imaginary part is
  considerably smaller than that for the real part. The imaginary part is odd
  in Re$[z]$ as well as in Im$[z]$.}
\end{figure}

\begin{figure}[!ht]
  \unitlength1.0cm
  \begin{center}
  \begin{picture}(3.0,2.0)
  \put(-5.,-5.){
  \psfig{file=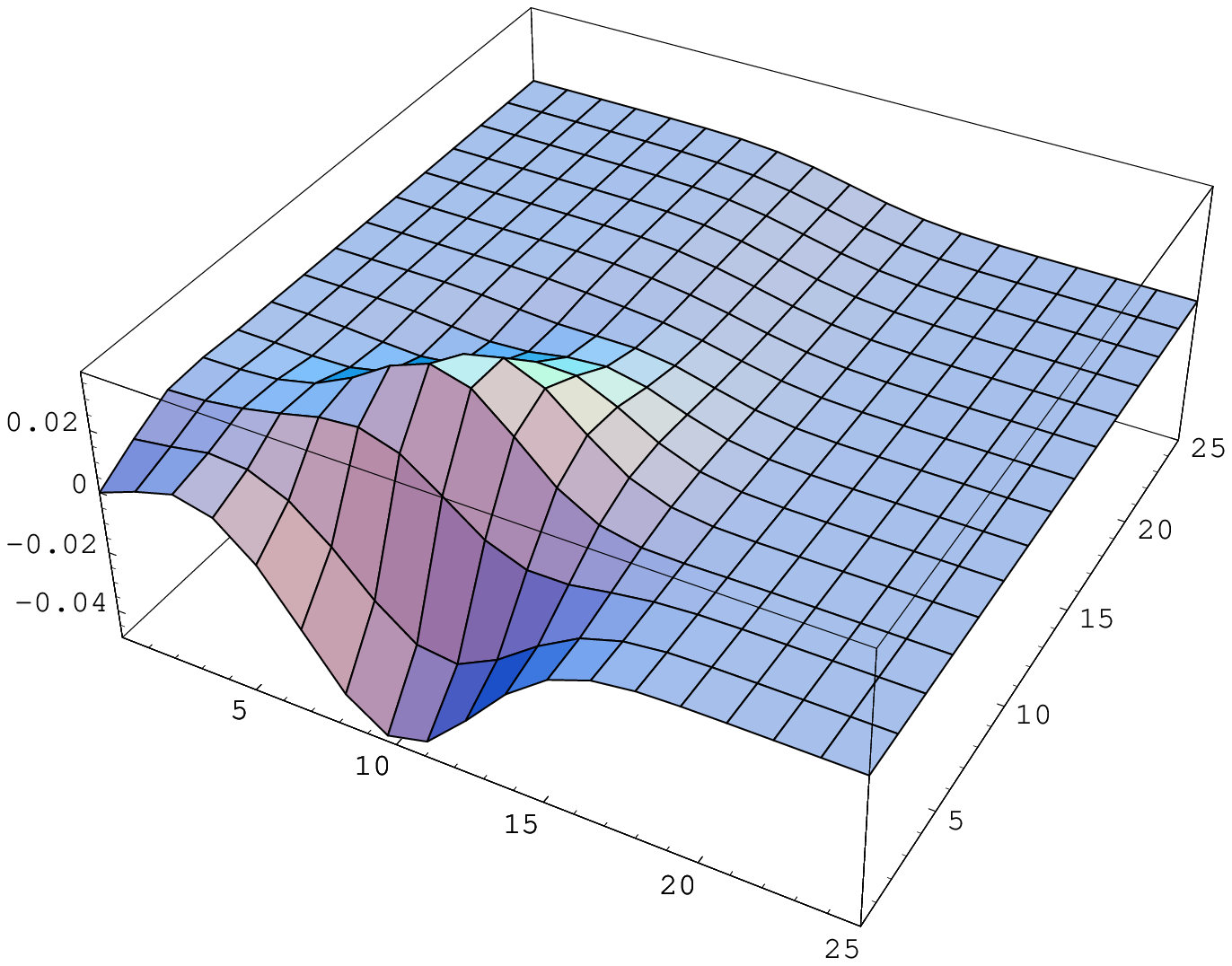,clip=,width=10cm}}
  \put(-2.3,-4.5){\bf\large Re$[z]\Sigma V$} 
  \put(5,-1.4){\bf\large Im$[z]\Sigma V$}
  \put(-6.8,2.4){\bf \LARGE $\frac{{\rm Re}[\rho^{N_f=1}_{\nu =0}(z,z^*,m;\mu)]}{\Sigma^2V^2}$}
  \put(-5.,-14.){
  \psfig{file=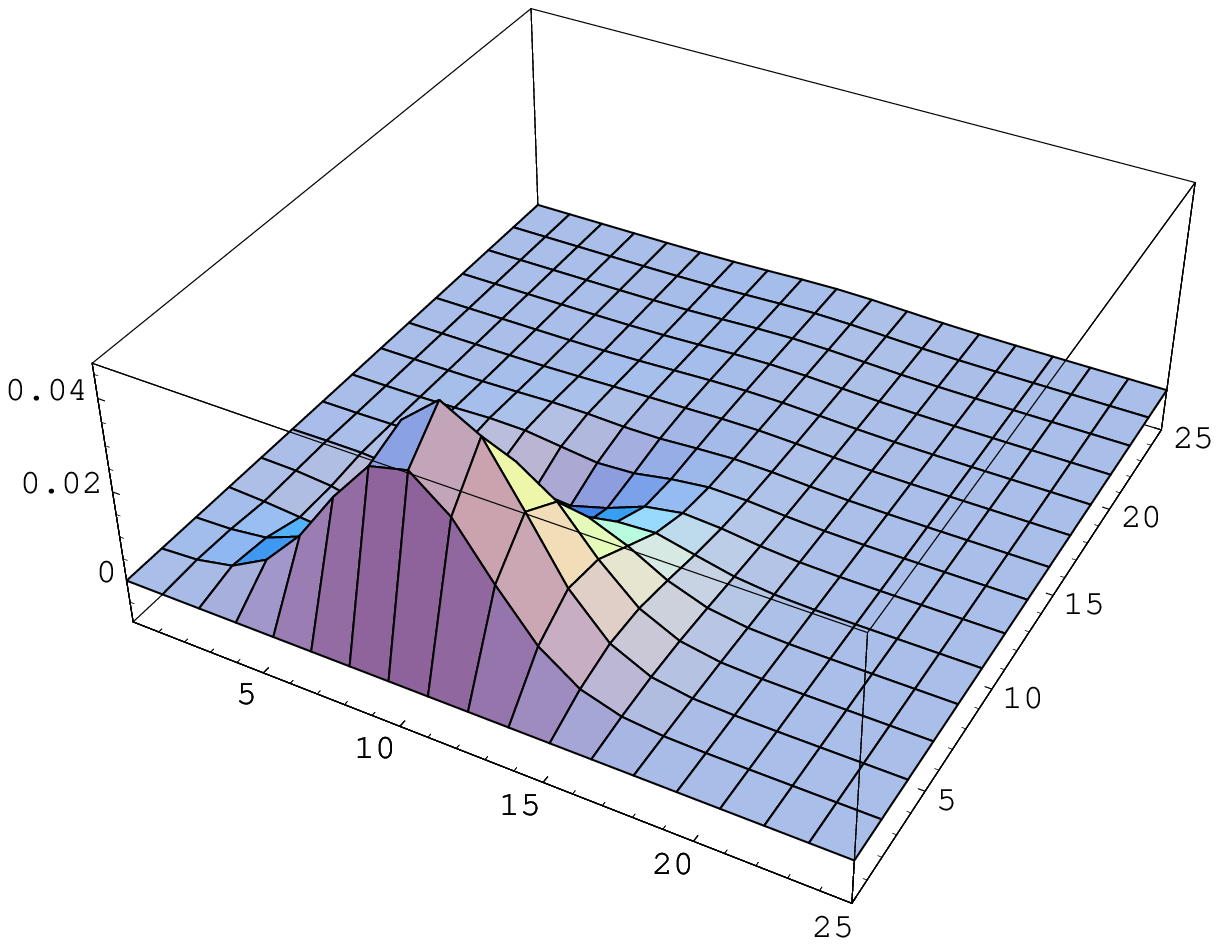,clip=,width=10cm}}
  \put(-2.3,-13.5){\bf\large Re$[z]\Sigma V$}
  \put(5,-10.4){\bf\large Im$[z]\Sigma V$}
  \put(-6.8,-6.6){\bf \LARGE $-\frac{{\rm Im}[\rho^{N_f=1}_{\nu=0}(z,z^*,m;\mu)]}{\Sigma^2V^2}$}
  \end{picture}
  \vspace{14cm}
  \end{center}
\caption{\label{fig:Rerhoa1m10mu01B} Real and imaginary parts 
of the eigenvalue density for one 
  flavor of mass $mV\Sigma=5$ and chemical potential
  $\mu F_\pi \sqrt V=2.5$. 
With these values the real part of the density deviates
  substantially from the quenched density. 
  Note that the real part of the density changes sign at $z=m$.
  In this case the scale of the imaginary part is comparable to
  that of the real part.}
\end{figure}

\begin{figure}[ht]
  \unitlength1.0cm
  \begin{center}
  \begin{picture}(3.0,2.0)
  \put(-5.,-5.){
  \psfig{file=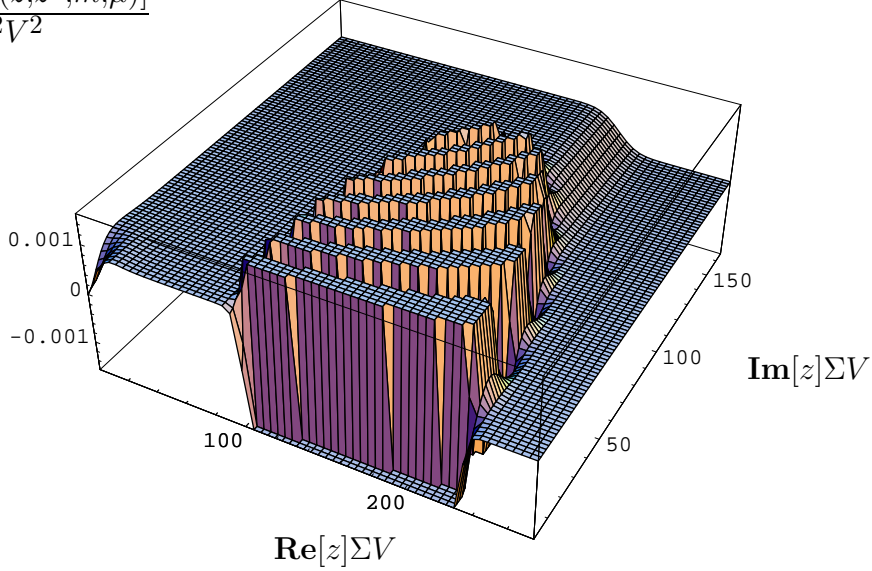,clip=,width=10cm}}
  \put(-1.3,-4.8){\bf\large Re$[z]\Sigma V$}
  \put(5,-2.4){\bf\large Im$[z]\Sigma V$}
  \put(-6.8,2.3){\bf \LARGE $\frac{{\rm Re}[\rho^{N_f=1}_{\nu=0}(z,z^*,m;\mu)]}{\Sigma^2V^2}$}
  \end{picture}
  \vspace{5cm}
  \end{center}
  \caption{\label{fig:ReRhoa1m100mu10}  
  The real part of the eigenvalue density for one flavor of mass
  $m\Sigma V=100$ and chemical potential $\mu F_\pi \sqrt{V}=10$.
  The severe sign
  problem manifest itself in the strongly oscillating region of the
  eigenvalue density which has amplitudes of the order $\exp(10)$.
  The full size of the peaks has been clipped for better illustration.}
\end{figure}

\begin{figure}[ht]
  \unitlength1.0cm
  \begin{center}
  \begin{picture}(-1.0,2.0)
  \put(-5.,-5.){
  \psfig{file=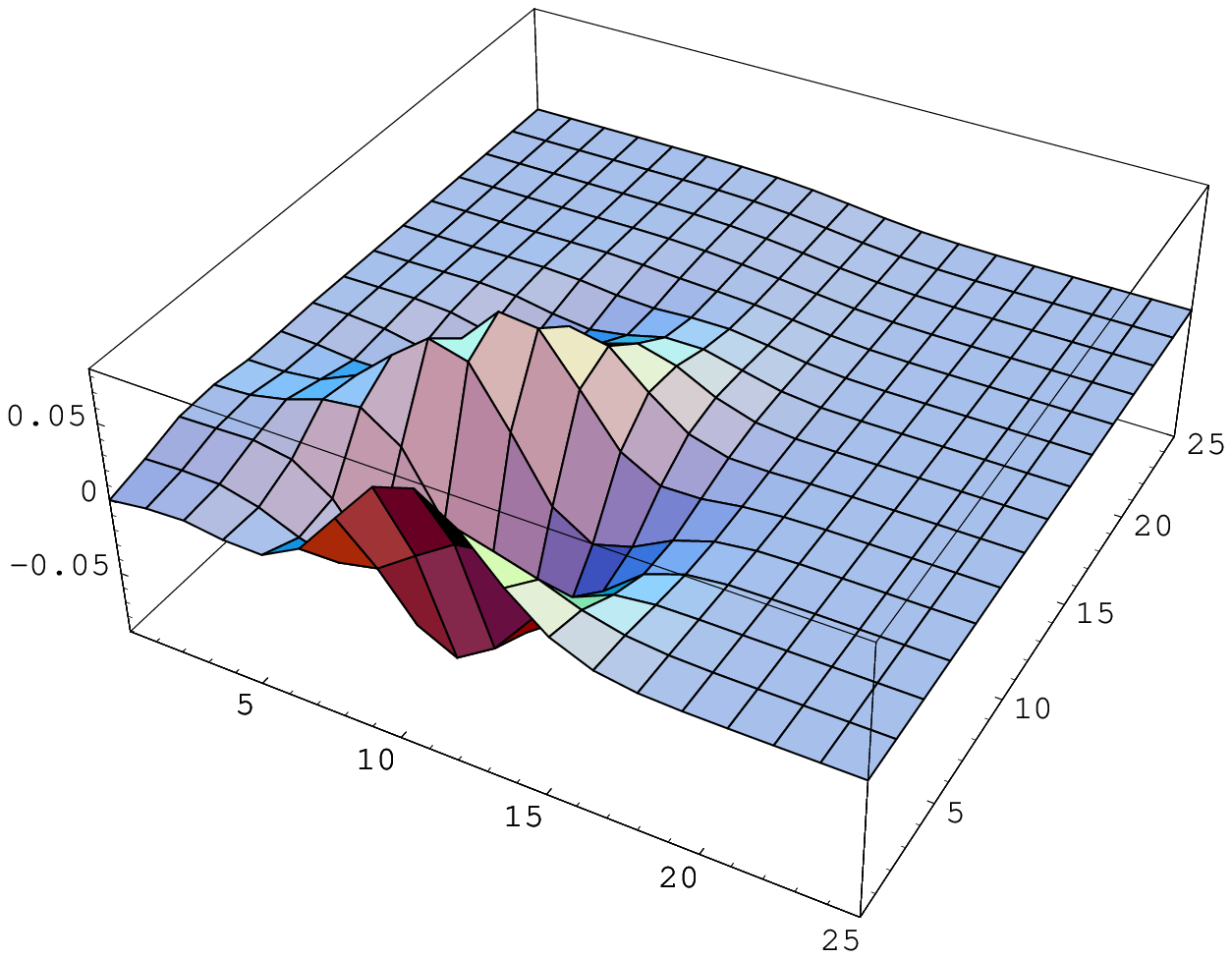,clip=,width=10cm}}
  \put(-1.3,-4.9){\bf\large Re$[z]\Sigma V$}
  \put(5,-1.4){\bf\large Im$[z]\Sigma V$}
  \put(-8.0,2.1){\bf \LARGE $\frac{{\rm Re}[\rho^{N_f=2}_{\nu=0}(z,m_1,m_2;\mu)]}{\Sigma^2V^2}$}
  \put(-5.,-14.){
  \psfig{file=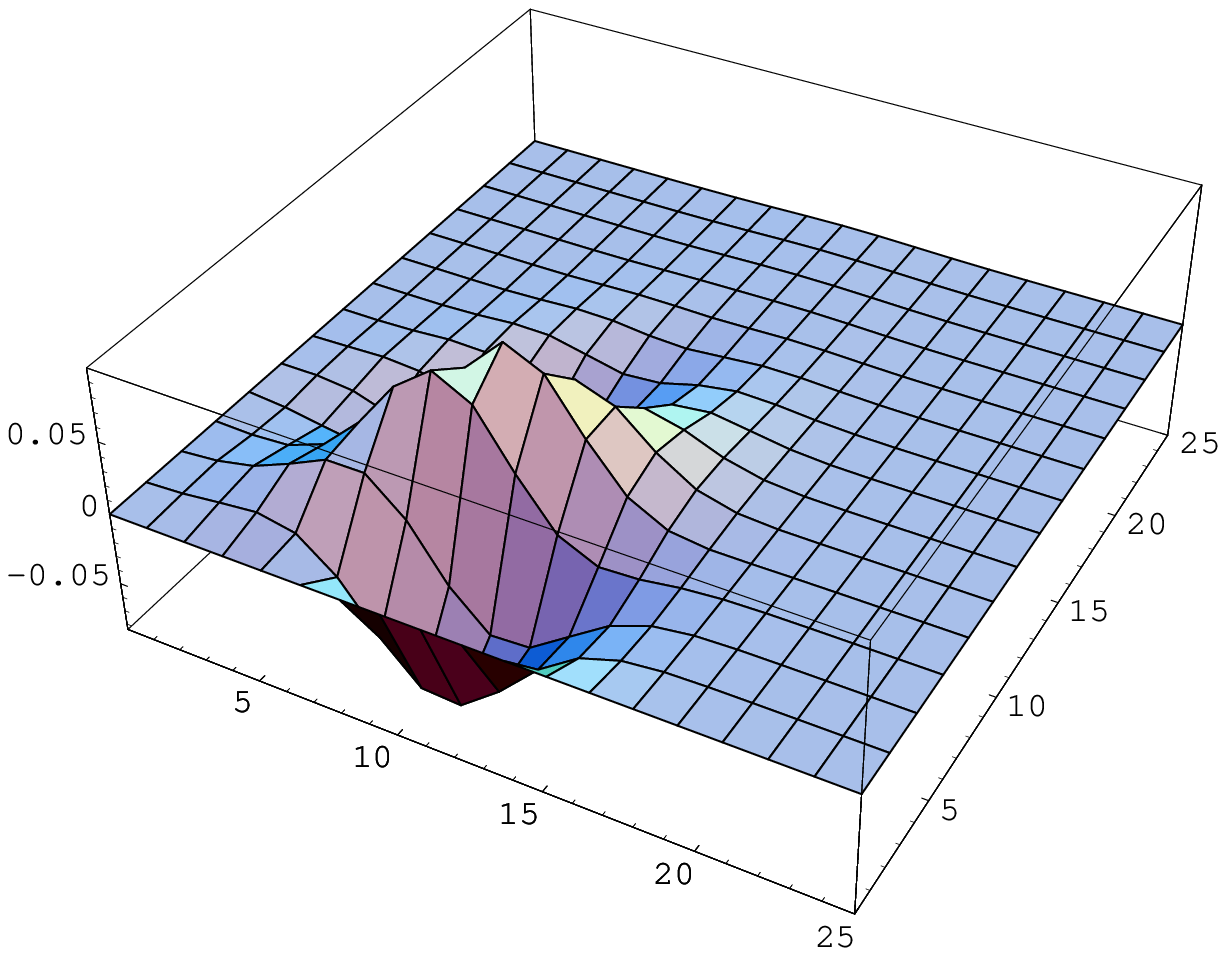,clip=,width=10cm}}
  \put(-1.3,-13.8){\bf\large Re$[z]\Sigma V$}
  \put(5,-10.4){\bf\large Im$[z]\Sigma V$}
  \put(-8.0,-6.7){\bf \LARGE $-\frac{{\rm Im}[\rho^{N_f=2}_{\nu=0}(z,m_1,m_2;\mu)]}{\Sigma^2V^2}$}
  \end{picture}
  \vspace{14cm}
  \end{center}
  \caption{\label{fig:ReRhoa2m5mu2p5} 
    Real and imaginary parts 
    of the spectral density with two flavors of mass
    $m_1 V\Sigma=m_2 V\Sigma=5$ and
    chemical potential $\mu F_\pi \sqrt V=2.5$. The density is
    neither real nor positive and the fluctuations are an order of magnitude
    larger than the quenched density at the same value of $\mu$. 
      The spectral density
    has a zero at $z=m_1$. The  imaginary part of the spectral density
    is anti-symmetric about the real and imaginary axis.}   
\end{figure}

The microscopic eigenvalue density with one dynamical quark of mass $m$ and
with baryon chemical potential $\mu$ is given by \cite{O}
\be
\rho^{N_f=1}_{\nu=0}(z,z^*,m;\mu) 
& = & \frac{|z|^2\Sigma^4V^3}{2\pi \mu^2F_\pi^2}\mbox{e}^{-2\mu^2 F_\pi^2 V} 
K_0\left(\frac{|z|^2\Sigma^2 V}{4\mu^2F_\pi^2}\right)
\mbox{e}^{-\frac{(z^2+z^{*\,2})\Sigma^2 V}{8\mu^2F_\pi^2}}\frac{\left|
\begin{array}{cc}   
I_0(m\Sigma V) & I_0(z\Sigma V)  \nn\\
\tKz(m,z^*;\mu) & \tKz(z,z^*;\mu)    
\end{array}\right|}{ I_0(m\Sigma V) }\nn\\
& = & 
\rho^{N_f=0}_{\nu = 0}(z,z^*;\mu)\left(1-\frac{I_0(z\Sigma V)\tKz(m,z^*;\mu)}
{I_0(m\Sigma V)\tKz(z,z^*;\mu)}\right).
\label{rhoNf1}
\ee
This result is shown in Fig. \ref{fig:Rerhoa1m10mu01} 
for $\mu F_\pi\sqrt V= 0.1$ 
and $mV\Sigma =10$, and in Fig. \ref{fig:Rerhoa1m10mu01B} for 
$\mu F_\pi\sqrt V=2.5$ and
$mV\Sigma=5$.
It is our first example of a spectral density that is
complex due to the sign problem.
%which causes a sign problem in numerical simulations.

Since the unquenched spectral density is proportional to the quenched
spectral density its support cannot go beyond the support of the 
quenched spectral density. In particular for $\mu^2F_\pi^2V \gg 1$ 
the unquenched spectral density is also confined inside the strip
given by (\ref{zstrip}). As we can observe from
Fig. \ref{fig:ReRhoa1m100mu10} 
the asymptotic behavior is much more complicated than in the quenched case.
An asymptotic form of the integral (\ref{zn=1}) was derived in 
\cite{gernotSpectra}.
For $\mu^2F_\pi^2V \gg 1$ and $(x+m)\Sigma/(4\mu^2F_\pi^2)<1$, 
it is well approximated by 
\be
 \int_0^1 dt \, t e^{-2\mu^2F_\pi^2V t^2} I_0((x-iy)\Sigma Vt)
 I_0(m\Sigma Vt) 
&\approx &
 \int_0^\infty dt \, t  e^{-2\mu^2F_\pi^2V t^2} I_0((x-iy)\Sigma Vt) 
 I_0(m\Sigma Vt)\nn \\
&=& \frac 1{4\mu^2F_\pi^2V}e^{\frac{((x-iy)^2+m^2)\Sigma^2
     V}{8\mu^2F_\pi^2}}  
I_0\left(\frac{m(x-iy)\Sigma^2 V}{4\mu^2F_\pi^2}\right).
\ee
This approximation breaks down if $(x+m)\Sigma/(4\mu^2F_\pi^2)>1$ because
then the saddle point of the $t$-integration will be outside the interval
$[0,1]$. The exponential functions in the asymptotic behavior of the 
Bessel functions no longer compensate each other in the additional
contribution to the unquenched spectra density.
Instead of a plateau for $N_f =0$ we find an oscillatory contribution
with an amplitude that increases exponentially with the volume.
For the total spectral
density a plateau is still visible in the region where the quenched
contribution dominates. As an illustration we show in Fig.
\ref{fig:ReRhoa1m100mu10} the real part of the eigenvalue density
for $\mu^2F^2_\pi V = 100$ and $mV\Sigma =100$. The period of the oscillations
is of the order of the level spacing at $\mu=0$ whereas the amplitude
is of the order $\exp(10)$ for our choice of parameters.

\subsubsection{The spectral density for two flavors}

For two flavors of mass $m_1$ and $m_2$ at nonzero baryon chemical
potential the density of eigenvalues is given by \cite{O}
\be
\rho^{N_f=2}_{\nu = 0}(z,z^*,m_1,m_2;\mu)
& = & \frac{|z|^2\Sigma^4V^3}{2\pi\mu^2F_\pi^2}\mbox{e}^{-2\mu^2 F_\pi^2 V} 
K_0\left(\frac{|z|^2\Sigma^2 V}{4\mu^2F_\pi^2}\right)
\mbox{e}^{-\frac{(z^2+z^{*\,2})\Sigma^2 V}{8\mu^2F_\pi^2}}\hspace{3cm}\nn\\
&& \hspace{1cm}\times
\frac{\left|\begin{array}{ccc} 
I_0(m_1\Sigma V) & I_0(m_2\Sigma V) & I_0(z\Sigma V)   \\   
 m_1\Sigma V I_1(m_1\Sigma V) & m_2\Sigma V 
I_1(m_2\Sigma V) & z\Sigma V I_1(z\Sigma V)  \\
  \tKz(m_1,z^*;\mu) & \tKz(m_2,z^*;\mu)& \tKz(z,z^*;\mu) 
\end{array}\right|}{\left|\begin{array}{cc} 
I_0(m_1\Sigma V) & m_1\Sigma V I_1(m_1\Sigma V) \\   
I_0(m_2\Sigma V) & m_2\Sigma V I_1(m_2\Sigma V) 
\end{array}\right|}.
\label{rhoNf2}
\ee
In Fig. \ref{fig:ReRhoa2m5mu2p5} we show the real and imaginary parts of
the spectral density for $\mu F_\pi\sqrt V=2.5$ and 
$m_1\Sigma V=m_2\Sigma V=5$. The spectral density
is neither real nor positive.

\subsubsection{The spectral density with one pair of conjugate flavors}

%\vspace*{2cm}
\begin{figure}[ht]
  \unitlength1.0cm
  \begin{center}
  \begin{picture}(3.0,2.0)
  \put(-5.,-5.){
  \psfig{file=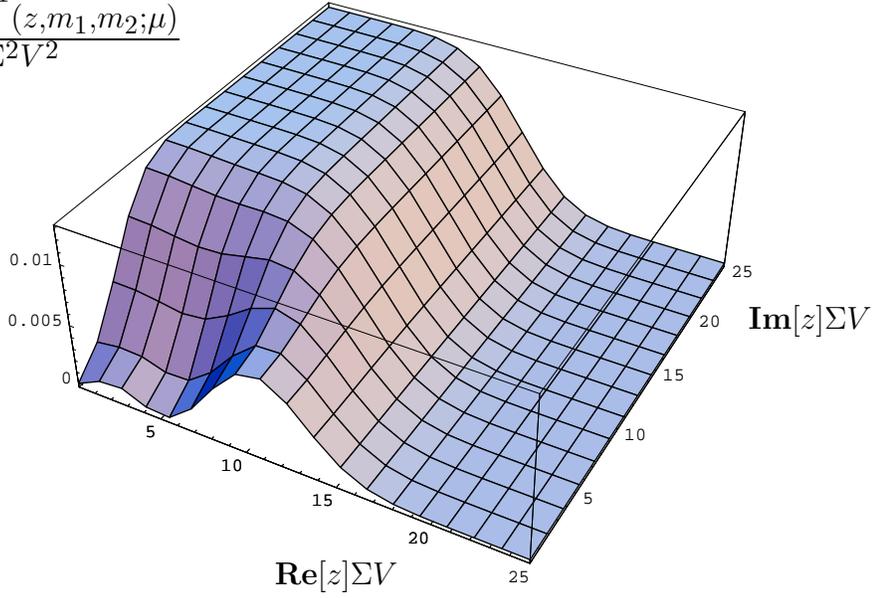,clip=,width=10cm}}
  \put(-1.3,-4.8){\bf\large Re$[z]\Sigma V$}
  \put(5,-1.4){\bf\large Im$[z]\Sigma V$}
  \put(-6.8,2.3){\bf \LARGE $\frac{\rho^{N_f=0,n=1}_{\nu=0}(z,m_1,m_2;\mu)}{\Sigma^2V^2}$}
  \end{picture}
  \vspace{5cm}
  \end{center}
  \caption{
  \label{fig:densn1m5mu2p5} The spectral density with one pair of conjugate 
  flavors of mass $m_1 V\Sigma=m_2 V \Sigma=5$
 and chemical potential $\mu F_\pi \sqrt V=2.5$.
  It is real and positive with a zero at $z=m_1$.  
Since the masses are real and equal this corresponds to a nonzero isospin
  chemical potential.}     
\end{figure}

The microscopic spectral density of the Dirac operator in QCD with one pair of
conjugate quarks (with mass $m$ and $m^*$) is found to be (see (\ref{rhol})
and  
\cite{AV})  
\be
\rho^{n=1}_{\nu = 0}(z,m,m^*;\mu) & = & 
\frac{|z|^2\Sigma^4V^3}{2\pi \mu^2F_\pi^2} \mbox{e}^{-2\mu^2 F_\pi^2 V} 
K_0\left(\frac{|z|^2\Sigma^2 V}{4\mu^2F_\pi^2}\right)
\mbox{e}^{-\frac{(z^2+z^{*\,2})\Sigma^2 V}{8\mu^2F_\pi^2}}     
\frac{ \left|\begin{array}{cc} \tKz(m,m^*;\mu) & \tKz(m,z^*;\mu) \nn \\
                            \tKz(z,m^*;\mu) & \tKz(z,z^*;\mu)
                           \end{array}\right|}
{\tKz(m,m^*;\mu)}\label{rhoQ1} \\
 & = &
 \rho^{N_f=0}_{\nu=0}(z;\mu)\left(1-\frac{\tKz(z,m^*;\mu)\tKz(m,z^*;\mu)}
{\tKz(m,m^*;\mu)\tKz(z,z^*;\mu)}\right).
\label{rhon1}
\ee
From this expression it follows that the density is real, positive,
and has a zero at $z=m$. This is illustrated by
Fig.  \ref{fig:densn1m5mu2p5}, where we plot the spectral density for
$m V\Sigma=5$ and $\mu F_\pi\sqrt V = 2.5$.

Since the quark mass, $m$, is real, the determinant of the conjugate
quark is identical to the determinant of a quark with a chemical potential of
the same magnitude but with opposite sign, cf. (\ref{conj-iso}). That is, 
the partition function in this case is equivalent to a theory with
two degenerate flavors and nonzero isospin chemical potential, $\mu_I$.  
This version of QCD does not have a sign problem \cite{AKW} as has been 
explored by lattice QCD simulations \cite{Kogut-Sinclair}. 

\subsubsection{The density with one ordinary flavor and one pair of
  conjugate flavors}

%\hspace*{3cm}\vspace*{3cm}
\begin{figure}[!ht]
  \unitlength1.0cm
  \vspace{2cm}
  \begin{center}
  \begin{picture}(3.0,2.0)
  \put(-5.,-5.){
  \psfig{file=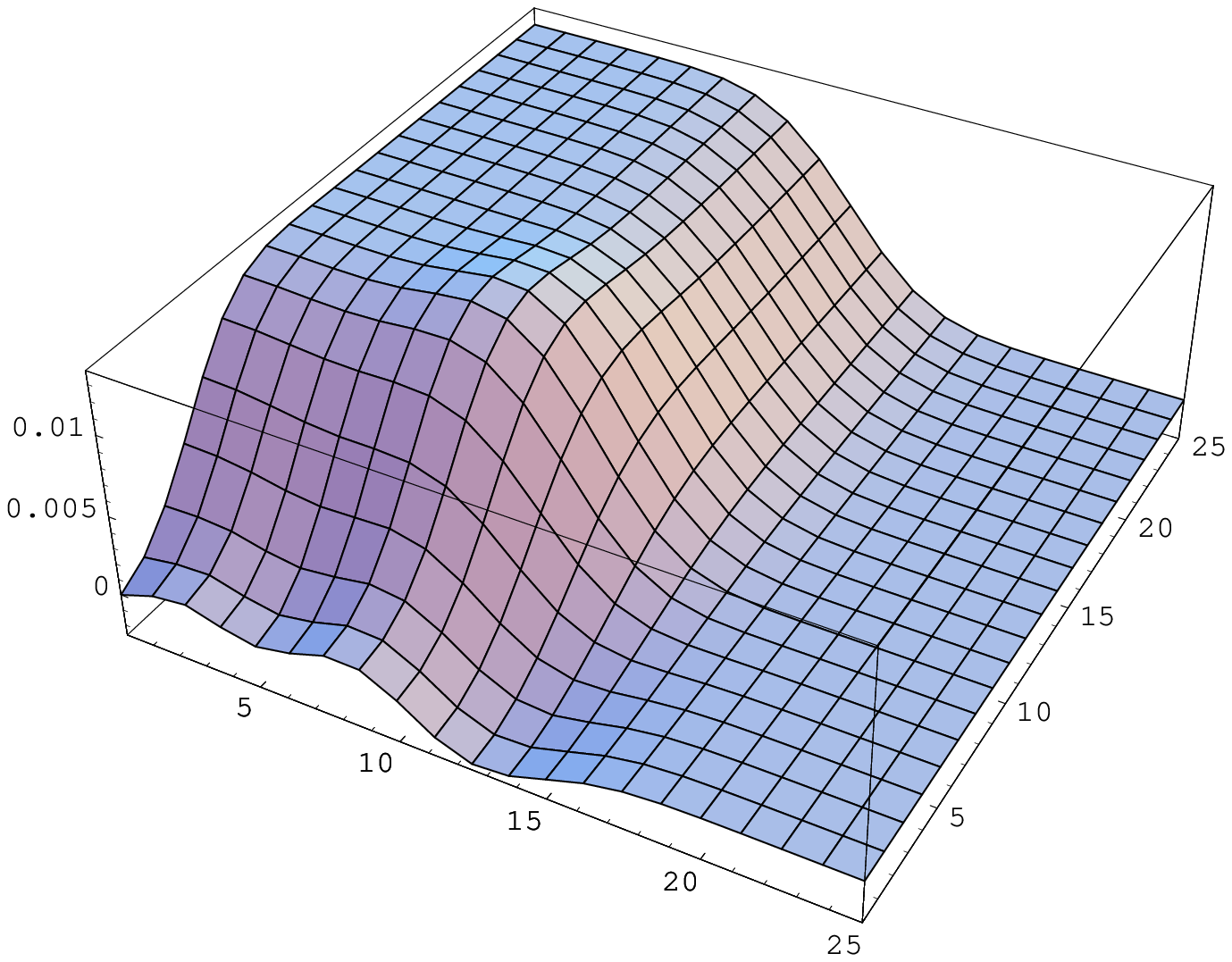,clip=,width=10cm}}
  \put(-1.3,-4.8){\bf\large Re$[z]\Sigma V$}
  \put(5,-1.4){\bf\large Im$[z]\Sigma V$}
  \put(-6.8,3.3){\bf \LARGE $\frac{{\rm Re}[\rho^{N_f=1,n=1}_{\nu=0}(z,m,m,m_s;\mu)]}{\Sigma^2V^2}$}
  \end{picture}
  \vspace{5cm}
  \end{center}
\caption{\label{fig:nf1n1} 
The real part of the density in a theory with 3 quarks. The masses are
$m_u V\Sigma=m_d V \Sigma=5$ and $m_s V\Sigma=10$. The chemical potentials
are $\mu_s=\mu_I=\mu=2.5/(F_\pi\sqrt V)$. Note that the density bounces
off the real axis at $z=m_u$ and changes sign at $z=m_s$ as expected.}
\end{figure}

As a final example we write out the microscopic eigenvalue density in QCD
with 3 light flavors with real masses $m_u=m_d=m$ and $m_s$  
and chemical potentials $\mu_I=\mu_s=\mu$. This theory has a sign problem
due to the extra quark that isn't matched by a conjugate quark.
%even though $\mu_B=0$.
The phase diagram as a function of $\mu_I$ and $\mu_s$
was determined in \cite{KT}. The eigenvalue density is (\ref{rhoNfandl})
\be
\rho^{N_f=1,n=1}_{\nu =0}(z,z^*,m_s,m,m;\mu)
& = &\frac{|z|^2\Sigma^4V^3}{2\pi\mu^2F_\pi^2} \mbox{e}^{-2\mu^2 F_\pi^2 V} 
K_0\left(\frac{|z|^2\Sigma^2 V}{4\mu^2F_\pi^2}\right)
\mbox{e}^{-\frac{(z^2+z^{*\,2})\Sigma^2 V}{8\mu^2F_\pi^2}} \hspace{3cm}\nn \\
&&\hspace{1cm}
\times\frac{\left|\begin{array}{ccc} 
I_0(m\Sigma V)  &  I_0(m_s\Sigma V) & I_0(z\Sigma V) \\
\tKz(m,m;\mu) & \tKz(m_s,m;\mu) & \tKz(z,m;\mu)   \\
\tKz(m,z^*;\mu) & \tKz(m_s,z^*;\mu) & \tKz(z,z^*;\mu)
\end{array}\right|}{\left|\begin{array}{cc} 
I_0(m\Sigma V) & I_0(m_s\Sigma V) \\   
\tKz(m,m;\mu) &   \tKz(m_s,m;\mu)
\end{array}\right|}.
\label{rhoNf1n1}
\ee
For $z=m$ the two top rows are identical and the first and third
column are identical in the $3\times3$ determinant. Hence the density is
zero at $z=m$ but does not change sign.  While for $z=m_s$ the first
and second column in the $3\times3$ determinant are identical, so the
density changes sign at $z=m_s$.  A plot of this density is shown in
Fig. \ref{fig:nf1n1}.

\section{Conclusions}
\label{conclu}

We have analyzed the spectrum of the QCD Dirac operator at nonzero
baryon chemical potential. In
the microscopic limit this spectrum is uniquely determined by the
global symmetries of the QCD partition function. This is true both
for the quenched and the unquenched theory. In both cases the 
spectral density of the  Dirac operator
can only be obtained after the introduction of a complex conjugate
pair of valence quarks resulting in a nontrivial baryon charge matrix 
and Goldstone modes with nonzero baryon number. The 
microscopic limit of the generating function for the QCD Dirac spectrum
can therefore be represented 
in two different ways. First, as an integral over the Goldstone manifold
which is determined by the pattern of symmetry breaking. Second, as the
large-$N$ limit of a partition function of
an ensemble of non-Hermitian matrices with the 
symmetries of the QCD partition function. In recent work, 
the Dirac spectrum was derived directly from the random matrix model
by means of complex orthogonal polynomials.  
In this paper we have obtained 
the Dirac spectrum from the replica limit of the Toda lattice equation. 
This explains that the spectral density factorizes into the product
of a fermionic partition function and the partition function for 
one conjugate pair of bosonic valence quarks. The bosonic factor does
not depend on the dynamical quarks and therefore goes beyond what
is implied by the Toda lattice structure.
This arises as a consequence of a singularity in the partition functions
that contain a pair of  conjugate bosonic 
quarks with the same mass. This
singularity  persists even for finite size matrices. In fact, it is present
for an ensemble of  $2\times 2$ random matrices. Also the Toda
lattice structure is valid for finite size matrices, and one can easily 
verify that the correct spectral density is obtained for an ensemble
of $2\times 2$ matrices. 

As a new result, we have also obtained  explicit expressions
for the spectral density when the absolute value of the fermion
determinant is present in the QCD partition function. In this case
a direct check from lattice simulations is possible for all values of the
scaled chemical potential, as has already been done in the fully quenched
case. 

The calculation of the unquenched 
microscopic spectral density has been  a highly nontrivial test of the 
Toda lattice approach. We expect that it will be equally successful
for all remaining 
$\beta =2$ cases. For example, the method could be applied to the
calculation of GUE $S$-matrix fluctuations and parametric
correlations \cite{jv-unp}
as well as to other
non-Hermitian random matrix ensembles \cite{david}.
The extension of this approach to Dyson index
 $\beta =1$ and $\beta =4$ is still an open problem.

The  spectral density
can be obtained from the replica limit of a Painlev\'e equation that has been 
derived from fermionic partition  functions only.
On the other hand, if a family of partition functions 
satisfies a Painlev\'e equation, a Toda 
lattice equation can be derived from the B\"acklund transformations of the
Painlev\'e system. Therefore, all systems in the same universality class
have the same microscopic limit for the fermionic partition functions, 
the bosonic partition functions and the microscopic spectral density.

Another important difference between Hermitian and non-Hermitian
Gaussian random matrix theories
is that the probability density of non-Hermitian theories
do not factorize into 
the joint eigenvalue density and the distribution of the eigenvectors.
The joint eigenvalue density is only obtained after a nontrivial 
integration over the similarity transformations that diagonalize
the non-Hermitian matrices. This results in the appearance of
a nontrivial universal factor in the joint eigenvalue distribution, in our
case a modified $K$-Bessel function. The appearance of this function is
a direct consequence of the chiral symmetry of the problem. 

 The microscopic spectral density derived in this paper
is valid for small temperatures
 and all values of the dimensionless parameters $\mu^2F_\pi^2V$ and
 $m\Sigma V$.
 Within this range of parameters, where the sign problem develops and becomes
 severe, the effect on the spectral density can be followed
 analytically.
 A first manifestation of the sign problem is that the spectral density in
 the complex eigenvalue plane is no longer real. This
occurs  already in a parameter range where lattice simulations might
be feasible.  For larger values of $\mu^2F_\pi^2V$ the analytical
 expressions for the eigenvalue density can serve as tests for numerical
 methods that apply to the nonperturbative regime of QCD at nonzero
 baryon chemical potential. In the limit where $\mu^2F_\pi^2V\gg1$
 the spectral ``density'' shows oscillations with a period of the order
 of the level spacing and an amplitude that diverges exponentially
 with $\mu^2F_\pi^2V$. This behavior, which is not present in the quenched
 case, should be responsible for the breaking of chiral symmetry. This
 issue will be addressed in a future publication.

\vspace{1cm}

\noindent
{\bf Acknowledgments:} 
We wish to thank the INT in Seattle
where this work was conceived.
Dominique Toublan is thanked for collaboration on the graded integrals
and Poul Henrik Damgaard for discussions on orthogonal polynomials. 
G.A. was supported in part by a Heisenberg Fellowship,
J.C.O. was supported by U.S. NSF grant PHY 01-39929 and
 U.S. DOE grant DE-FC02-01ER41180,
and J.J.M.V. was supported in part by U.S. DOE grant DE-FG-88ER40388.

\vspace*{1cm}\noindent

%%%%%%%%%%%%%%%%%%%%%%%%%%%%%%%%%%%%%%%%%%%%%%%%%%%%%%%%%%%%%%%%%%%%%%%%%%%%%

\renewcommand{\thesection}{Appendix \Alph{section}}
\setcounter{section}{0}

\section{Some details on two flavor and general partition functions}
\label{app:Z}

For completeness we collect here the formulas for the remaining two 
partition functions with two flavors.
As was observed in \cite{AV,AP}, partition functions that do not
mix  quarks or conjugate quarks in either the numerator or denominator 
can be written solely as determinants of polynomials and Cauchy
transforms. We find
\be
\frac{{\cal Z}_N^{N_f=1,\,N_b=1}(x|y^*;\mu)}{{\cal Z}_N^{N_f=0}} &\equiv& 
\left\langle\ \left(\frac{x}{y^*}\right)^\nu\prod_{j=1}^N \frac{(z_j^2 - x^2)}{(z_j^{*\,2} - y^{*\,2})}
\ \right\rangle_{N_f=0} \ \ =\ \frac{-1}{r_{N-1}}\left(\frac{x}{y^*}\right)^\nu
\left|\begin{array}{ll}
h_{N-1}(y^*) & h_{N}(y^*)\\
p_{N-1}(x) & p_{N}(x)\\
\end{array}
\right|,
\label{hpZ}
\ee
and
\be
\frac{{\cal Z}_N^{N_b=2}(y_1,y_2;\mu)}{{\cal Z}_N^{N_f=0}} &\equiv& 
\left\langle\ \frac{1}{(y_1y_2)^\nu}\prod_{j=1}^N \frac{1}{(z_j^2 - y_1^2)(z_j^{2} - y_2^{2})}
\ \right\rangle_{N_f=0} \nn \\
 &=& \frac{1}{r_{N-1}r_{N-2}(y_1^2-y_2^2)(y_1y_2)^\nu}
\left|\begin{array}{ll}
h_{N-2}(y_1) & h_{N-1}(y_1)\\
h_{N-2}(y_2) & h_{N-1}(y_2)\\
\end{array}
\right|.
\label{hhZ}
\ee
All remaining combinations of bosons and fermions and their conjugates 
can be obtained from the given ones by complex conjugation.

Next we wish to comment on the relation between these formulas and in
particular on the additional pole terms which occur in the two kernels
containing Cauchy transforms, equations (\ref{Adef}) and (\ref{Ndef}).
For random matrix models with real eigenvalues the kernel of the polynomials
(\ref{Kdef}) satisfies the Christoffel-Darboux identity,
\be
K_N(s,t)&=& \frac{1}{r_{N-1}}
\frac{p_N(s)p_{N-1}(t)-p_N(t)p_{N-1}(s)}{s^2-t^2}\ ,\ \ s,t\in
\mbox{\bf R}\ .
\label{CD}
\ee
For orthogonal polynomials in the complex plane (\ref{OPdef}) 
this relation is generally not satisfied so that (\ref{KpreZ}) 
and (\ref{ppZ}) are different. In the Hermitian limit 
$\mu\to0$ they are equal, due to the relation (\ref{CD}).

The Eqs.~(\ref{AZ}) and (\ref{hhZ}) or 
(\ref{NZ}) and (\ref{hpZ}) can be related if the following
Christoffel-Darboux identities are valid:
\be
{\cal A}_{N-2}(s,t)&=& \frac{1}{r_{N-2}}
\frac{h_{N-1}(s)h_{N-2}(t)-h_{N-1}(t)h_{N-2}(s)}{s^2-t^2}\ ,\ \ s,t\in
\mbox{\bf R}\nn\\
{\cal N}_{N-1}(s,t)&=& \frac{1}{r_{N-1}}
\frac{h_N(s)p_{N-1}(t)-p_N(t)h_{N-1}(s)}{s^2-t^2}\ ,\ \ s,t\in
\mbox{\bf R}\  .
\label{CDCauchy}
\ee
This is the case for Hermitian random matrix models, but in general there
are no such Christoffel-Darboux identities for non-Hermitian random
matrix theories. This identity follows from the observation that 
in general the orthogonal 
polynomials and their Cauchy transforms obey the same three-step recursion
relation. 
Only the recursion involving the lowest Cauchy transform $h_{k=0}(y)$ is
different from that involving the polynomial $p_{k=0}(y)$, which results
in the extra integral $Q(s,t)$ over poles in (\ref{Adef}),
or the single pole $1/(s^2-t^2)$ in (\ref{Ndef}), respectively.
Thus in the Hermitian limit (\ref{AZ}) and (\ref{hhZ}), and 
(\ref{NZ}) and (\ref{hpZ}) become equal due to their  Christoffel-Darboux
identities (\ref{CDCauchy}).
This explains the presence of pole terms in (\ref{AZ}) and (\ref{NZ}).
In (\ref{Ndef}) this term can also be understood differently, 
as it ensures the correct
normalization of the partition functions (\ref{NZ})
to unity at equal arguments.

Let us now turn to more general partition functions. We will briefly explain
here how our generating function (\ref{ZNNfn-1}) follows from 
the results of \cite{BII} (generalized to the chiral ensemble).
There the following expectation value of complex eigenvalues is computed as
\be
\label{B1}
\left\langle\ \prod_{j=1}^N \frac{\prod_{f=1}^{N_f}(m_f^2-z_j^2 )}{
(z_j^2 - y^2)(z_j^{*\,2} - x^{*\,2})}
\ \right\rangle_{N_f=0} 
&=&\frac{(-1)^{N_f}}{r_{N-1}}\frac{
\left|
\begin{array}{lll}
p_{N-1}(m_1) & \ldots\ p_{N-1}(m_{N_f}) & h_{N-1}(x)^*\\
~~~~\vdots & \ddots~~~~\vdots & ~~~~\vdots\\
p_{N+N_f-2}(m_1) & \ldots\ p_{N+N_f-2}(m_{N_f}) & h_{N+N_f-2}(x)^*\\
{\cal N}_{N+N_f-2}(m_1,y) &  \ldots\ {\cal N}_{N+N_f-2}(m_{N_f},y)&
{\cal A}_{N+N_f-2}(x^*,y)\\
\end{array}
\right|
}{
\left|
\begin{array}{lll}
p_{0}(m_1) & \ldots\ p_{0}(m_{N_f}) & h_{0}(x)^*\\
~~~~\vdots & \ddots~~~~\vdots & ~~~~\vdots\\
p_{N_f-2}(m_1) & \ldots\ p_{N_f-2}(m_{N_f}) & h_{N_f-2}(x)^*\\
0&\ldots\ 0& p_{0}(x)^*\\
{\cal N}_{N_f-2}(m_1,y) &  \ldots\ {\cal N}_{N_f-2}(m_{N_f},y)&
{\cal A}_{N_f-2}(x^*,y)\\
\end{array}
\right|}. \nn \\
\label{before}
\ee  
%Notice that for $N_f=1$ the last row in the determinant of the denominator is
%absent. 
In order to simplify the denominator 
the determinant can be expanded with respect to the 
next to last row. The remaining determinant only contains polynomials 
$p_k(m_i)$ and the kernel ${\cal N}_{N_f-2}(m_i,y)$. 
The invariance of determinants under adding and subtracting rows and columns 
can be used to first eliminate all sums in the kernels 
${\cal N}_{N_f-2}(m_i,y)$. Then all polynomials can be reduced to 
monomials, leading to the following Vandermonde like determinant
\be
\left|
\begin{array}{lll}
1&\ldots & 1\\
m_1^2 & \ldots& m_{N_f}^2 \\
~~\vdots & \ddots & ~~\vdots\\
m_1^{2(N_f-2)} & \ldots& m_{N_f}^{2(N_f-2)}\\
\frac{1}{m_1^2-y^2}&  \ldots& \frac{1}{m_{N_f}^2-y^2}\\
\end{array}
\right| =
\left|
\begin{array}{lll}
\frac{m_1^2}{m_1^2-y^2}&\ldots & \frac{m_{N_f}^2}{m_{N_f}^2-y^2}\\
\frac{m_1^4}{m_1^2-y^2}& \ldots& \frac{ m_{N_f}^4}{m_{N_f}^2-y^2}\\
~~~~\vdots & \ddots & ~~~~\vdots\\
\frac{m_1^{2(N_f-1)}}{m_1^2-y^2}& \ldots& 
\frac{m_{N_f}^{2(N_f-1)}}{m_{N_f}^2-y^2}\\
\frac{1}{m_1^2-y^2}&  \ldots& \frac{1}{m_{N_f}^2-y^2}\\
\end{array}
\right| =\ (-1)^{N_f-1} 
\frac{\Delta_{N_f}(\{m_f^2\})}{\prod_{f=1}^{N_f}(m_f^2-y^2)}.
\ee
Here we have again made use of invariance properties of the determinant 
before arriving at the desired result. Substituting this in
(\ref{before}) we obtain  (\ref{ZNNfn-1}).

%%%%%%%%%%%%%%%%%%%%%%%%%%%%%%%%%%%%%%%%%%%%%%%%%%%%%%%%%%%%%%%%%%%%%%%%%%%%%%

\section{The microscopic eigenvalue density with conjugate quarks} 
\label{app:todaConjugate}

In this appendix we derive the microscopic eigenvalue density of the Dirac
operator in a theory with conjugate fermionic quarks. 
%\vspace{2mm}
Even if some or all of the fermionic flavors are conjugate quarks the Toda
lattice equation holds. In the microscopic limit the partition functions
\be
{\cal Z}_\nu^{N_f,l,n}(\{m_f\},\{m_c\},\{m_c^*\},z,z^*;\mu) & = &
\int[{\rm d}A]_\nu\prod_{f=1}^{N_f}\det(D(\mu)+m_f)\nn\\ &&\times
\prod_{c=1}^l|\det(D(\mu)+m_c)|^2
|\det(D(\mu)+z)|^{2n}\mbox{e}^{-S_{\rm YM}},   
\label{ZBaryonIsoSpin}
\ee
satisfy the Toda lattice equation
\be  
\label{TodaBaryonIsoSpin} 
 && \delta_z\delta_{z^*} \log Z_\nu^{N_f,l,n}(\{m_f\},\{m_c\},\{m_c^*\},z,z^*;\mu) 
\nn\\
& = & \frac{\pi n}2 (zz^*)^2 
\frac{Z_\nu^{N_f,l,n+1}(\{m_f\},\{m_c\},\{m_c^*\},z,z^*;\mu)
      Z_\nu^{N_f,l,n-1}(\{m_f\},\{m_c\},\{m_c^*\},z,z^*;\mu)}
{[Z_\nu^{N_f,l,n}(\{m_f\},\{m_c\},\{m_c^*\},z,z^*;\mu)]^2}.
\ee
This follows by a direct extension of the argument given in section 2.3.2 of
\cite{SplitVerb2}.   
In order to obtain the density with $l$ pairs of conjugate quarks
we consider the replica limit of the Toda lattice equation 
(\ref{TodaBaryonIsoSpin}) with $N_f=0$  
\be 
\rho^{l}_\nu(z,z^*,\{m_c\},\{m_c^*\};\mu) 
&=& 
\frac{zz^*}2\frac{Z_\nu^{l,n=1}(\{m_c\},\{m_c^*\},z,z^*;\mu)  
Z_\nu^{l,n=-1}(\{m_c\},\{m_c^*\}|z,z^*;\mu)}  
{[Z_\nu^{l}(\{m_c\},\{m_c^*\};\mu)]^2}.  
\ee  
Using (\ref{ZNfn=-1}) we find the spectral density for a theory with
$l$ pairs of conjugate quarks
\be
\rho^{l}_\nu(z,z^*,\{m_c\},\{m_c^*\};\mu) 
& = & \frac{zz^*}2
\prod_{c=1}^{l}|z^2-m_c^2|^2 Z_{n=-1}(z,z^*;\mu)
\frac{Z_\nu^{l,n=1}(\{m_c\},\{m_c^*\},z,z^*;\mu)}
{Z_\nu^{l}(\{m_c\},\{m_c^*\};\mu)}
\label{rhol}.\nn \\
\ee
This density is positive and real for all masses and chemical
potentials. This is fully consistent with having a real and positive measure
in (\ref{defdens}).  We have explicitly checked that this eigenvalue
density is consistent with what we obtain using the orthogonal
polynomial method as described in \cite{AV}.

Finally, we give the most general result we have obtained using this
method, namely the spectral density for a theory with $N_f+l$ quarks and
$l$ conjugate quarks,
\be\label{rhoNfandl}
&& \rho_\nu^{N_f,l}(z,z^*,\{m_f\},\{m_c\},\{m_c^*\};\mu) \\ 
& = & \frac{zz^*}2
\prod_{f=1}^{N_f}(z^2-m_f^2)\prod_{c=1}^{l}|z^2-m_c^2|^2 Z_{n=-1}(z,z^*;\mu)
\frac{Z_\nu^{N_f,l,n=1}(\{m_f\},\{m_c\},\{m_c^*\},z,z^*;\mu)}
{Z_\nu^{N_f,l}(\{m_f\},\{m_c\},\{m_c^*\};\mu)}.
\nn 
\ee

%%%%%%%%%%%%%%%%%%%%%%%%%%%%%%%%%%%%%%%%%%%%%%%%%%%%%%%%%%%%%%%%%%%%%%%


\begin{thebibliography}{9}

\bibitem{Barbour} I. Barbour et al., Nucl. Phys. { B 275} (1986) 296.

\bibitem{owe}
P.~de Forcrand and O.~Philipsen,
%``The QCD phase diagram for small densities from imaginary chemical
%potential,''
Nucl.\ Phys.\ B { 642} (2002) 290 
[arXiv:hep-lat/0205016].
%%CITATION = HEP-LAT 0205016;%%

\bibitem{maria}
M.~D'Elia and M.P.~Lombardo,
%``Finite density QCD via imaginary chemical potential,''
Phys.\ Rev.\ D { 67} (2003) 014505 
[arXiv:hep-lat/0209146].
%%CITATION = HEP-LAT 0209146;%%

%\cite{Allton:2002zi}
\bibitem{allton}
C.R.~Allton {\it et al.},
%``The QCD thermal phase transition in the presence of a small chemical
%potential,''
Phys.\ Rev.\ D { 66} (2002) 074507 
[arXiv:hep-lat/0204010].
%%CITATION = HEP-LAT 0204010;%%

\bibitem{frit}
F.~Karsch,
%``Lattice QCD at non-zero chemical potential and the resonance gas model,''
Prog.\ Theor.\ Phys.\ Suppl.\  { 153} (2004) 106 
[arXiv:hep-lat/0401031].
%%CITATION = HEP-LAT 0401031;%%


%\cite{Gocksch:1987ha}
\bibitem{Gocksch}
A.~Gocksch,
%``On Lattice QCD At Finite Density,''
Phys.\ Rev.\ D { 37} (1988) 1014.
%%CITATION = PHRVA,D37,1014;%%

\bibitem{misha}M.A. Stephanov, Phys. Rev. Lett. { 76} (1996) 4472
[arXiv:hep-lat/9604003].

\bibitem{SV}E.V. Shuryak and J.J.M. Verbaarschot, 
Nucl. Phys. A { 560} (1993) 306 [arXiv:hep-th/9212088]; 
J.J.M. Verbaarschot, Phys. Rev. Lett. { 72} (1994) 2531 [arXiv:hep-th/9401059].


%\cite{Verbaarschot:1995yi}
\bibitem{Vplb}
J.J.M.~Verbaarschot,
%``Universal scaling of the valence quark mass dependence of the chiral
%condensate,''
Phys.\ Lett.\ B { 368} (1996) 137 
[arXiv:hep-ph/9509369].
%%CITATION = HEP-PH 9509369;%%


\bibitem{GL}         J. Gasser and H. Leutwyler, 
                      Ann. Phys. { 158} (1984) 142;         %OK
                      Nucl. Phys. B { 250} (1985) 465;      %OK
                      H. Leutwyler, 
                      Ann. Phys. { 235} (1994) 165.         %OK


\bibitem{Weinberg} S.~Weinberg, Phys.~Rev. { 166} (1968) 1568.
\bibitem{KST}         J.B. Kogut, M.A. Stephanov and D. Toublan, 
                      % On two-color QCD with baryon chemical potential
                      % hep-ph/9906346%%CITATION =  HEP-PH 9906346;%%
                      Phys. Lett. B { 464} (1999) 183  [arXiv:hep-ph/9906346].
          %OK

\bibitem{KSTVZ} J.B. Kogut, M.A. Stephanov, D. Toublan, J.J.M. Verbaarschot
and A. Zhitnitsky, 
Nucl.\ Phys.\ B { 582} (2000) 477 [arXiv:hep-ph/0001171].
%%CITATION = HEP-PH 0001171;%%

\bibitem{dominique-JV}D. Toublan and J.J.M. Verbaarschot,
  Int. J. Mod. Phys. B { 15} (2001) 1404 [arXiv:hep-th/0001110].

\bibitem{eff}  
D.T.~Son and M.A.~Stephanov,
%``QCD at finite isospin density,''
Phys.\ Rev.\ Lett.\  { 86} (2001) 592 [arXiv:hep-ph/0005225]; 
%%CITATION = HEP-PH 0005225;%% 
%Phys.~Atom.~Nucl. { 64} (2001) 834;
%Yad.~Fiz. { 64} (2001) 899. 
K.~Splittorff, D.T.~Son and M.A.~Stephanov, 
%``QCD-like theories at finite baryon and isospin density,''
Phys.\ Rev.\ D { 64} (2001) 016003 [arXiv:hep-ph/0012274];
%%CITATION = HEP-PH 0012274;%% 
J.B.~Kogut and D.~Toublan,
%``QCD at small non-zero quark chemical potentials,''
Phys.\ Rev. D { 64} (2001) 034007  [arXiv:hep-ph/0103271]; 
K. Splittorff, D. Toublan and J.J.M. Verbaarschot,
  Nucl.~Phys. B { 620} (2002) 290 [arXiv:hep-ph/0108040]; 
%%CITATION = HEP-PH 0108040;%% 
 Nucl. Phys. B { 639} (2002) 524  [arXiv:hep-ph/0204076]; 
%%CITATION = HEP-PH 0204076;%% 
 J. Wirstam, J.T. Lenaghan  and K. Splittorff, Phys. Rev. D { 67} (2003)
 034021 [arXiv:hep-ph/0210447]; %%CITATION = HEP-PH 0210447;%% 
J.T. Lenaghan, F. Sannino and K. Splittorff, Phys. Rev. D { 65} (2002)
 054002  [arXiv::hep-ph/0107099]. %%CITATION = HEP-PH 0107099;%%

\bibitem{LS}H. Leutwyler and A. Smilga, Phys. Rev. D { 46} (1992) 5607.

\bibitem{GLeps}  J. Gasser and H. Leutwyler, 
Phys. Lett. B { 188} (1987) 477.
%Thermodynamics of chiral symmetry

\bibitem{OTV}         J.C. Osborn, D. Toublan and J.J.M. Verbaarschot,
                      % From chiral Random Matrix Theory to chiral
                      % Perturbation Theory vz     % hep-th/98
                       Nucl. Phys. B { 540} (1999) 317
                      [arXiv:hep-th/9806110].          %OK

\bibitem{DOTV}      P.H. Damgaard, J.C. Osborn, D. Toublan and  
J.J.M. Verbaarschot,
                      % hep-th/9811212
                       Nucl. Phys. B { 547} (1999) 305
                       [arXiv:hep-th/9811212].       %OK

\bibitem{AD4}  G.~Akemann and P.H.~Damgaard,
%hep-th/0311171
Phys. Lett. { B 583} (2004) 199  [arXiv:hep-th/0311171].

\bibitem{EA}
S.F. Edwards and P.W. Anderson, J. Phys. F{ 5} (1975) 965.

\bibitem{replica}
%\bibitem{EJ}
S.F. Edwards and R.C. Jones, J. Phys. A { 9}  (1976) 1595;
%\bibitem{VZR}
J.J.M. Verbaarschot and M.R. Zirnbauer, Ann. Phys. {
158} (1984) 78;
J.J.M. Verbaarschot and M.R. Zirnbauer,
J. Phys. A { 18} (1985) 1093;
%\bibitem{KM}
A. Kamenev and M.~Mezard, J. Phys. A { 32} (1999) 4373
[arXiv:cond-mat/9901110];
%Wigner-Dyson statistics from the replica method
Phys. Rev. B { 60} (1999) 3944 [arXiv:cond-mat/9903001];
%Level correlations in disordered metals: The replica sigma model.
%\bibitem{lerner}
I.V. Yurkevich and I.V. Lerner,
Phys. Rev. B { 60} (1999) 3955  [arXiv:cond-mat/9903025];
%Nonperturbative results for level correlations from the replica nonlinear
%sigma model. 
%\bibitem{zirn}
M.R. Zirnbauer, arXiv:cond-mat/9903338.

\bibitem{kanzieper02}E. Kanzieper, Phys. Rev. Lett. { 89} (2002) 250201
 [arXiv:cond-mat/0207745].

\bibitem{kanzieper03}E. Kanzieper, to appear in {\it Progress in Field Theory
  Research}, Nova Science, New York 2004 
{ [arXiv:cond-mat/0312006]}.
% Exact replica treatment of non-Hermitean complex random matrices


\bibitem{SplitVerb1} K. Splittorff and J.J.M. Verbaarschot,
  Phys. Rev. Lett. { 90} (2003)   041601 
 [arXiv:cond-mat/0209594].

\bibitem{SplitVerb2} K. Splittorff and J.J.M. Verbaarschot,
  Nucl. Phys. B { 683} (2004) 467  [arXiv:hep-th/0310271].



\bibitem{Korepin} V.E. Korepin, N.M. Bogoliubov and A.G. Izergin, 
{\it Quantum inverse scattering method and correlation functions}, 
Cambridge Univ. Press, 1993.



\bibitem{ForresterBook} P. Forrester, {\sl Log-gases and Random matrices},
  Web Book - available at 
{\tt http://www.ms.unimelb.edu.au/$\sim$matpjf/matpjf.html}.


\bibitem{SplitVerb3} 
%\cite{Splittorff:2004ru}
K.~Splittorff and J.J.M.~Verbaarschot,
%``Supersymmetric quenching of the Toda lattice equation,''
Nucl.\ Phys.\ B { 695} (2004) 84 
[arXiv:hep-th/0402177].
%%CITATION = HEP-TH 0402177;%%



\bibitem{Efetov} K.B. Efetov, Phys. Rev. Lett. { 79}, 491 (1997);
                      Adv. Phys. { 32}, 53 (1983),                %OK
                     {\it Supersymmetry in disorder and chaos},
                     (Cambridge University Press, Cambridge, 1997).
                                %OK

\bibitem{Kharchev}
S.~Kharchev, A.~Marshakov, A.~Mironov, A.~Morozov and A.~Zabrodin,
%``Towards unified theory of 2-d gravity,''
Nucl.\ Phys.\ B { 380} (1992) 181 
[arXiv:hep-th/9201013].


\bibitem{ADII} 
G. Akemann and P. H. Damgaard,  Nucl. Phys. { B 576} (2000) 597 
[arXiv:hep-th/9910190];
H.W. Braden, A. Mironov and A. Morozov, Phys. Lett. { B 514} (2001) 293 
[arXiv:hep-th/0105169].  

\bibitem{AFV} G. Akemann, Y.V. Fyodorov and G. Vernizzi, 
Nucl. Phys. B 694 (2004) 59 [arXiv:hep-th/0404063].

\bibitem{O}  J. C. Osborn, Phys. Rev. Lett., in press [arXiv:hep-th/0403131].

\bibitem{francesco}
P.~Di Francesco, M.~Gaudin, C.~Itzykson and F.~Lesage,
%``Laughlin's wave functions, Coulomb gases and expansions of the
%discriminant,''
Int.\ J.\ Mod.\ Phys.\ A { 9} (1994) 4257 
[arXiv:hep-th/9401163].
%%CITATION = HEP-TH 9401163;%%

\bibitem{FKS}Y.V. Fyodorov, B.A. Khoruzhenko and H.-J. Sommers,
Phys. Rev. Lett. { 79} (1997) 557  [arXiv:cond-mat/9703152];
Ann. Inst. Henri Poincar\'e { 68} (1998) 449 [arXiv:chao-dyn/9802025].

\bibitem{gernotSpectra} G. Akemann, Phys. Rev. Lett. { 89} (2002) 072002
[arXiv:hep-th/0204068];  
 J. Phys. A { 36} (2003) 3363 
[arXiv:hep-th/0204246].

\bibitem{AV} G. Akemann and G. Vernizzi, Nucl. Phys. B { 660} (2003) 532
 [arXiv:hep-th/0212051]. 

%\cite{Bergere:2004cp}
\bibitem{BII}
M.C.~Bergere,
%``Biorthogonal Polynomials for Potentials 
%of two Variables and External Sources
%at the Denominator,''
arXiv:hep-th/0404126.
%%CITATION = HEP-TH 0404126;%%

\bibitem{AP} G. Akemann and  A. Pottier, 
J. Phys. A 37 (2004) L453 [arXiv:math-ph/0404068].





\bibitem{tilomar}
H.~Markum, R.~Pullirsch and T.~Wettig,
%``Non-Hermitian random matrix theory and lattice QCD with chemical
%potential,''
Phys.\ Rev.\ Lett.\  { 83} (1999) 484 
[arXiv:hep-lat/9906020].
%%CITATION = HEP-LAT 9906020;%%

\bibitem{hands}
E.~Bittner, S.~Hands, H.~Markum and R.~Pullirsch,
%``Quantum chaos in supersymmetric QCD at finite density,''
Prog.\ Theor.\ Phys.\ Suppl.\  { 153} (2004) 295 
[arXiv:hep-lat/0402015].
%%CITATION = HEP-LAT 0402015;%%

\bibitem{GT} G. Akemann and T. Wettig, Phys. Rev. Lett. { 92} (2004)
  102002 [arXiv:hep-lat/0308003]. 


\bibitem{bittner}
E.~Bittner, M.P.~Lombardo, H.~Markum and R.~Pullirsch,
%``Dirac and Gorkov spectra in two color QCD with chemical potential,''
Nucl.\ Phys.\ Proc.\ Suppl.\  { 94} (2001) 445 
[arXiv:hep-lat/0010018];
%%CITATION = HEP-LAT 0010018;%%
E.~Bittner, M.P.~Lombardo, H.~Markum and R.~Pullirsch,
%``Lowest eigenvalues of the Dirac operator for two color QCD at nonzero
%chemical potential,''
Nucl.\ Phys.\ Proc.\ Suppl.\  { 106} (2002) 468 
[arXiv:hep-lat/0110048].
%%CITATION = HEP-LAT 0110048;%%
\bibitem{gbittner}
G.~Akemann, E.~Bittner, M.P.~Lombardo, H.~Markum and R.~Pullirsch,
%``Density profiles of small Dirac operator eigenvalues for two color QCD at
%nonzero chemical potential compared to matrix models,''
arXiv:hep-lat/0409045.
%%CITATION = HEP-LAT 0409045;%%





\bibitem{Kogut-Sinclair} J.B.~Kogut and D.K.~Sinclair, Phys. Rev. D { 66}
  (2002) 034505 [arXiv:hep-lat/0202028].
% {\tt hep-lat/0202028} Lattice QCD at finite isospin density at zero
% and finite temperature. 

\bibitem{Nakamura} A. Nakamura and T. Takaishi,
Nucl. Phys. { B} (Proc. Suppl.) { 129 \& 130} (2004) 629
 [arXiv:hep-lat/0310052].


\bibitem{Girko}       V.L. Girko, 
                      {\it Theory of random determinants}
                      (Kluwer Academic Publishers, Dordrecht, 1990).  %OK


\bibitem{minn04}
K.~Splittorff and J.J.M.~Verbaarschot,
%``QCD Dirac spectra and the Toda lattice,''
arXiv:hep-th/0408107.
%%CITATION = HEP-TH 0408107;%%


%\bibitem{Kan}
%E.~Kanzieper, in {\it Progress in Field Theory}, 
%Nova Science Publishers, 2004, 
%``Exact replica treatment of non-Hermitean complex random matrices,''
%arXiv:cond-mat/0312006.
%%CITATION = COND-MAT 0312006;%%
%cond-mat/0312006.




\bibitem{JSV-1}
A.~D.~Jackson, M.~K.~Sener and J.~J.~M.~Verbaarschot,
%``Universality near zero virtuality,''
Nucl.\ Phys.\ B 479 (1996) 707
[arXiv:hep-ph/9602225].
%%CITATION = HEP-PH 9602225;%%

\bibitem{ADMN}     
G. Akemann, P. Damgaard, U. Magnea and S. Nishigaki,
Nucl. Phys. B  487 [FS] (1997) 721 [arXiv:hep-th/9609174].

\bibitem{Guhr-tilo}
T.~Guhr and T.~Wettig,
%``Universal spectral correlations of the Dirac operator at finite
%temperature,''
Nucl.\ Phys.\ B 506 (1997) 589
[arXiv:hep-th/9704055].
%%CITATION = HEP-TH 9704055;%%

\bibitem{JSV-2}
A.~D.~Jackson, M.~K.~Sener and J.~J.~M.~Verbaarschot,
%``Universality of correlation functions in random matrix models of {QCD},''
Nucl.\ Phys.\ B 506 (1997) 612
[arXiv:hep-th/9704056].
%%CITATION = HEP-TH 9704056;%%

\bibitem{DN}    
 P.H. Damgaard and S.M. Nishigaki, Nucl. Phys. { B 518} (1998)
495 [arXiv:hep-th/9711023].


%\cite{Kanzieper:1998ti}
\bibitem{FK} E.~Kanzieper and V.~Freilikher,  in
NATO ASI, Series C (Math. and Phys. Sciences), Vol. 531, p. 165 - 211,
Kluwer, Dordrecht, 1999
%``Spectra of large random matrices: A method of study,''
[arXiv:cond-mat/9809365].
%%CITATION = COND-MAT 9809365;%%


\bibitem{DV}
D. Dalmazi and J.J.M. Verbaarschot, Nucl. Phys. B { 592} (2001) 419 
[arXiv:hep-th/0005229].
% The Replica Limit of Unitary Matrix Integrals, hep-th/0005229.

\bibitem{GF}
G. Akemann and  Y.V. Fyodorov, Nucl. Phys. B
{664} (2003) 457 [arXiv:hep-th/0304095].  


%\cite{Halasz:1996jg}
\bibitem{HJV}
M.A.~Halasz, A.~D.~Jackson and J.J.M.~Verbaarschot,
%``Yang-Lee zeros of a random matrix model for QCD at finite density,''
Phys.\ Lett.\ B { 395} (1997) 293 
[arXiv:hep-lat/9611008].
%%CITATION = HEP-LAT 9611008;%%



%G. Akemann and G. Vernizzi, Nucl. Phys. { B631} (2002) 471, 
%\bibitem{KV}    A.B.J. Kuijlaars and  M. Vanlessen,  
%Com. Math. Phys. { 243} (2003) 163 [arXiv:math-ph/0305044].


\bibitem{AD} G. Akemann  and P.H. Damgaard,  
Phys. Lett. { B 432} (1998) 390 [arXiv:hep-th/9802174].


 %\cite{Fyodorov:2002md}
\bibitem{Fyodor-ch}
Y.V.~Fyodorov and E.~Strahov,
%``On correlation functions of characteristic polynomials for chiral  Gaussian
%unitary ensemble,''
Nucl.\ Phys.\ B { 647} (2002) 581 
[arXiv:hep-th/0205215];
%%CITATION = HEP-TH 0205215;%%
Y.V. Fyodorov and G. Akemann, JETP Lett. 77 (2003) 438
[arXiv:cond-mat/0210647].
  
\bibitem{PZJ} P. Zinn-Justin, Commun. Math. Phys. { 194} (1998) 631
[arXiv:cond-mat/9705044].

\bibitem{Mehta}M.L. Mehta, {\it Random Matrices}, Academic Press, Second 
Edition, London 1991.


\bibitem{BI}
M.C.~Bergere,
%``Orthogonal polynomials for potentials of two variables with external
%sources,''
arXiv:hep-th/0311227.
%%CITATION = HEP-TH 0311227;%%

\bibitem{A02} G. Akemann, Phys. Lett. { B 547} (2002) 100
[arXiv:hep-th/0206086].

\bibitem{FS} Y.V. Fyodorov, B.A. Khoruzhenko and H.-J. Sommers, 
Phys. Lett. { A226} (1997) 46  [arXiv:cond-mat/9606173]; 
Y.V. Fyodorov and H.-J. Sommers, J.\ Math.\ Phys. { 38} (1997)
1918 [arXiv:cond-mat/9701037].


\bibitem{KSS} J.B. Kogut, M. Snow and M. Stone, Nucl. Phys. B {
    200} (1982) 211.  
 
\bibitem{Brower}  R.C. Brower, P. Rossi and  C-I. Tan,
                      % The external field problem for QCD
                      Nucl. Phys. B { 190} (1981) 699 ;        %OK
                      R.C. Brower and M. Nauenberg,
                      % Group integration for lattice gauge theory at
                      % large n and at small coupling
                      Nucl. Phys. B { 180} (1981) 221 .        %OK


\bibitem{jsv}
A.D. Jackson, M.K. Sener and J.J.M. Verbaarschot,
%``Finite volume partition functions and Itzykson-Zuber integrals,''
Phys.\ Lett.\ B { 387} (1996) 355 [arXiv:hep-th/9605183].
%%CITATION = HEP-TH 9605183;%

\bibitem{AKW}
M.~Alford, A.~Kapustin and F.~Wilczek,  Phys.~Rev. { D 59} (1999) 054502
[arXiv:hep-lat/9807039].


\bibitem{KT} J.B.~Kogut and D.~Toublan,
%``QCD at small non-zero quark chemical potentials,''
Phys.\ Rev. D { 64} (2001) 034007
[arXiv:hep-ph/0103271].


\bibitem{jv-unp}J.J.M. Verbaarchot, unpublished.

\bibitem{david} D. Bernard and A. LeClair, arXiv:cond-mat/0110649.
% A Classification of Non-Hermitian Random Matrices




\end{thebibliography}
\end{document}